\documentclass[british]{article}
\usepackage[T1]{fontenc}
\usepackage[utf8]{inputenc}
\usepackage{array}
\usepackage{multirow}
\usepackage{amsmath}
\usepackage{graphicx}
\usepackage[numbers]{natbib}

\makeatletter

\providecommand{\tabularnewline}{\\}
\newcommand{\lyxdot}{.}

\newcommand{\lyxaddress}[1]{
	\par {\raggedright #1
	\vspace{1.4em}
	\noindent\par}
}
\newenvironment{lyxlist}[1]
	{\begin{list}{}
		{\settowidth{\labelwidth}{#1}
		 \setlength{\leftmargin}{\labelwidth}
		 \addtolength{\leftmargin}{\labelsep}
		 }}
	{\end{list}}

\usepackage{hyperref}
\usepackage{breakurl}
\usepackage{threeparttable}

\makeatother

\usepackage{babel}
\begin{document}
\selectlanguage{british}
\title{Modelling population dynamics based on experimental trials with genetically
modified (RIDL) mosquitoes}
\author{Mario A. Natiello$^{\dagger}$ and Hernán G. Solari$^{\ddagger}$}
\maketitle

\lyxaddress{$^{\dagger}$Centre for Mathematical Sciences, Lund University, Box
118, S 221 00 LUND, Sweden. mario.natiello@math.lth.se (corresponding
author)}

\lyxaddress{$^{\ddagger}$Departamento de Física, FCEN-UBA and IFIBA-CONICET,
Argentina. solari@df.uba.ar}
\begin{abstract}
Recently, the RIDL-SIT technology has been field-tested for control
of \emph{Aedes aegypti}. The technique consists of releasing genetically
modified mosquitoes carrying a ``lethal gene''. In 2016 the World
Health Organization (WHO) and the Pan-American Health Organization
(PAHO) recommended to their constituent countries to test the new
technologies proposed to control \emph{Aedes aegypti} populations.
However, issues concerning effectiveness and ecological impact have
not been thoroughly studied so far. In order to study these issues,
we develop an ecological model. It presents interdependent dynamics
of mosquito populations and food in a homogeneous setting. Mosquito
populations are described in a stochastic compartmental setup, in
terms of reaction norms depending on the available food in the environment.
The development of the model allows us to indicate some critical biological
knowledge that is missing and could (should) be produced. Hybridisation
levels, release numbers during and after intervention and population
recovery time after the intervention as a function of intervention
duration and target are calculated under different hypotheses with
regard to the fitness of hybrids and compared with two field studies
of actual interventions. This minimal model should serve as a basis
for detailed models when the necessary information to construct them
is produced. For the time being, the model shows that nature will
not clean non-lethal introgressed genes.
\end{abstract}

\section{Introduction}

The use of the Sterile Insect Technique (SIT) has been proposed to
control populations of the mosquito \emph{Aedes aegypti (Ae ae.)}
because of public health reasons \citep{alph10}. \emph{Ae ae.} is
the main vector of several viruses such as the \emph{Flavivirus} producing
Yellow Fever, Dengue and Zika, as well as the \emph{Alphavirus} producing
Chikungunya. The idea is not new and follows the success obtained
with the fruit fly (\emph{Drosophila sp.}) in agricultural settings.

The early attempts to use the method with \emph{Ae ae.} sterilised
by radiation ended in failure \citep{morl62}. The evolution of SIT
by the introduction of genetically modified insects, GMI, incorporating
an autocidal gene \citep{gong05} (termed RIDL-SIT, Release of Insects
carrying a Dominant Lethal) was later considered to be a more promising
technique for mosquito control than irradiated mosquitoes \citep{bene03}.
The RIDL-SIT technique for \emph{Ae ae.} was developed recently and
communicated in a series of papers \citep{phuc07,harr11,harr12,lacr12,carv15,wins15}
(all references except the first one refer to actual field-tests).

The World Health Organization (WHO) recently promoted the realisation
of evaluations of this technique \citep{who16b} in the wake of the
2015 Zika pandemic in the Americas. More recently, the Food and Drug
Administration (FDA-USA) authorised the realisation of such evaluations
in the US \citep{fda16}. In correspondence with technological and
political developments, the Pan-American Health Organization (PAHO)
endorsed WHO's recommendations for its member states. It is apparent
that there exists a growing interest in performing tests of the RIDL-SIT
technique \emph{in the field}, despite the limited amount of available
information and the uncertainty of impact assessments. Nevertheless,
warnings were raised by scientists \citep{alag16} that foresaw the
ecological and social dimension of the problem, in correspondence
with the considerations of social scientists regarding technological
fixes \citep{reis13}.

In April 20-21, 2017 a scientific meeting was held at the Ministry
of Public Health of Argentina with the support of PAHO, addressing
the reasonability of conducting such experiments in the country. The
participants were the National health authorities, a PAHO representative,
a group of scientists proposing integrated control technologies (including
SIT), a group of scientists not involved in technological development
and two representatives of the foremost RIDL-SIT company. Despite
the presence of a scientist representing the team developing the RIDL-SIT
technique, some questions concerning impact assessment remained unanswered,
being nevertheless relevant for the scientific community and the society
as a whole, for example:
\begin{itemize}
\item Which lasting modifications of the environment will be produced by
these tests? 
\item Will the population of vectors previous to the environmental intervention
be re-established? If yes, how fast?
\end{itemize}
The goal of the present work is to address these and related issues
focusing on the technique promoted by the WHO \citep{who16b} concerning
the transgenic mosquito OX513A \citep{phuc07,harr11,harr12,lacr12,carv15,wins15}
(in the preliminary work \citep{phuc07} it is called LA513A) that
was preliminarily evaluated in \citep{fda16}. We will use the mathematical
methods of stochastic population dynamics, which should in general
be used before any intervention. We investigate the time-evolution
of a mosquito population as a result of the proposed interventions,
focusing on: 
\begin{itemize}
\item The hybridisation of the released mosquitoes with the local populations
as a result of genetic diffusion
\item The expected ratio between released males and local males necessary
to achieve different levels of control
\item The recovery time of the new, hybrid, populations when the intervention
ceases 
\item The reasons for the difficulties found in establishing such control
systems 
\end{itemize}
To achieve these goals we implement a stochastic model, able to deal
with small population numbers, based upon the available information.
Because of the scarcity of biological information concerning the released
mosquitoes, we have kept the model as general as possible, avoiding
the need to replace missing information with conjectures. We indicate
the most noticeable missing information in our critical review of
the data (see Section \ref{subsec:Lack-of-information}). It is also
worth mentioning that construction of conclusions concerning some
data available in the reports is occasionally inconsistent between
different reports and/or existing biological knowledge.

This manuscript is structured as follows: In Section \ref{sec:Bb}
we describe the biological background of the technique, including
a critical review of the available data, as well as the mathematical
methods supporting the simulation. Section \ref{sec:Results} displays
the outcome of the simulations and their relation to relevant biological
questions. In Section \ref{sec:field} results of this work are compared
with actual field observations, while Section \ref{sec:Conc} contains
a general discussion and concluding remarks.

\section{Biological background and mathematical methods\label{sec:Bb}}

\subsection{The RIDL-SIT technique}

The \emph{Sterile Insect Technique} (SIT) was introduced in the past
century for insect pest control. The goal is to release sterile males
in sufficient number to reduce (control) the population by reducing
the number of offspring. Sterility was originally achieved by radiation
and it has been successful as part of the measures to control the
Mediterranean fruit fly \citep{enke15} (see below), the \emph{screwworm
fly} \citep{hend05} and other insects regarded as pests (for a history
see \citep{klas05}). Hence, the idea of controlling populations of
mosquito vectors of viral diseases by sterilisation emerged clearly.
By 1962 \emph{Aedes aegypti} was considered a mosquito that could
be controlled with the technique. Field experiments were performed
by \citet{morl62} but despite overwhelming releases of sterilised
males in relation to the local mosquito population, no control was
made evident. In consideration of this failure the technique had to
be revised. \citet{bene03} discussed some conditions that a sterile
insect technique should satisfy to be successful; we quote:
\begin{quotation}
Apart from the obvious requisite of mating competitiveness, two criteria
must be satisfied by any sterilisation method: (1) sterility must
not be suppressed by complementation with any genotype present in
field populations. Any variation in the expression of the sterility
factor(s) following interaction with the large and heterogeneous genome
of the field population will quickly lead to selection for non-sensitive
females and fertile matings. (2) Absolute assurance of effectiveness
must result from the sterilisation method in the factory setting.
\end{quotation}
These authors also recommended that 
\begin{quotation}
a first step for the future deployment of transgenic technology is
to demonstrate the safety of release material in the absence of confounding
concerns about the spread of alleles and drive elements. This is possible
only if the released material is genetically sterile.
\end{quotation}
Several authors \citep{weid74,roge84,dobs02,alph10} have indicated
that density-dependent-effects can/will reduce the efficiency of SIT
programs to the point of making it non-viable.

To overcome in part these population (ecological) effects, the RIDL-SIT
technique was introduced. Transgenic mosquitoes were designed to have
a late acting dominant lethal gene, LG, \citep{phuc07} following
a technique developed for the fruit fly \citep{gong05}. The original
strains of modified mosquitoes were constructed using the Rockefeller
strain of \emph{Ae ae}., a strain that has been kept in captivity
for many years and presents adaptation of its life-cycle to the laboratory
\citep{Tejerina2009,Grech2010}, as well as manifestly larger sizes
and fertility than wild strains.

This work focuses on the technique developed upon the transgenic mosquito
OX513A specified in the previous Section. The proposed strategy is
to release solely male mosquitoes outnumbering the wild males and
thus strongly influencing the fecundation of wild females. The offspring
is expected to express the lethal gene and die before reaching fertile
age. Thus, the introduced mosquitoes would, after one generation,
disappear from the natural environment along with their offspring.
The expected outcome in a closed environment (no adult immigration
or emigration) is to control the original wild mosquito population.

The genetic engineering of the mosquitoes proceeds by inoculating
eggs with molecular constructs to ``induce tetracycline-repressible
dominant lethality in both males and females'' \citep{phuc07}, producing
four transgenic lines of modified mosquitoes which appear to be obtained
by modifications at a single site in the genome. Three of these lines
gave the desired mortality. One of these three presented increased
mortality near pupation and was selected for mass production because
of this desired attribute. According to \citep{phuc07}, a penetrance
of 93-97\% was achieved, meaning that at the level of phenotype a
3-7\% of the individuals carrying the gene do not manifest the early
mortality. We found no studies concerning whether the individuals
carrying the gene and arriving to adulthood transmit the trait of
not-manifesting-the-modified-gene. For the case of the Mediterranean
fruit fly, \citet{gong05} followed one generation ($n=4590$) concluding
that survival to adulthood despite the lethal gene is not an inheritable
trait. Such conclusion is inaccurate, a proper statement is that the
probability of survival being inheritable lies in the interval $[0,\,5.9\cdot10^{-4}]$
with $95\%$ confidence. The difference matters when this probability
is multiplied by millions of released individuals. Nevertheless, the
authors acknowledge that ``the possibility of biochemical resistance
to the lethal effector molecule remains a potential drawback to RIDL
that is not a significant issue for radiation-based sterilization''.

As part of the engineering process, a fluorescent marker was incorporated
to the same modified gene. This marker facilitates identification
of larvae of genetically modified mosquitoes.

The requisite of mating competitiveness \citep{bene03} of the modified
mosquitoes, is monitored by the quantity $C$, defined as the ratio
between offspring (lower case) to adult males (upper case) of modified
mosquitoes, $m/M$, normalised to the same ratio for the wild mosquitoes,
$w/W$: 
\[
C=\left(\frac{m}{M}\right)\left(\frac{w}{W}\right)^{-1}.
\]
Low estimates values of $C$ will have to be compensated by performing
larger releases. The estimation of $C$ performed in cage and field
experiments produced values of $0.059$ (Cayman Islands \citep{harr12})
and $0.031$ (Brazil, \citep{carv15}). We leave aside a previous
report for the Cayman Islands of $C=0.56$ \citep{harr11} obtained
in the context of trying to establish that mating of wild females
with modified males was actually possible.

An additional source of uncertainty is the lifespan of modified adults
in the wild. Mark, release and capture experiments were performed
twice. For \emph{Ae ae}., the ability to produce reliable data in
such experiments has been questioned. A large dispersion of results
is obtained when trying to estimate dispersal distances, i.e., dispersal
measurements appear to be greatly influenced by the measurement procedure
\citep{oter08,berg13}. The reported results appear not to escape
to this rule: survivals of $2$ days are reported in \citep{lacr12}
and of $5$ days in \citep{wins15}. In the first experiment the unmodified
laboratory strain was released as well showing a $2.1$ day survival
in average with no significant differences with respect to the modified
strain. The release of a control strain was not repeated in the second
experiment, probably because of the urban setting in which it was
performed. In contrast, the survival of modified mosquitoes tested
against a strain collected in Chiapas (Mexico) termed Latin Wild Type,
LWT, used as the basis of the FDA report \citep[Appendix F, p. 6-7]{fda16}
contains only experiments in the laboratory, but none in the field.
The experiment confirms the existence of significant differences in
phenotype between LWT and the GMI in terms of fertility and hatch
rate measured at the laboratory, the environment in which the GMI
was developed. We have not found data testing survival rates of modified
mosquitoes released in the field against the local wild population
under similar conditions among the reported literature (\citep{lacr12,wins15,fda16}).

\subsection{Mosquito, genetics and environment\label{subsec:Genetics-Environment}}

The populations of the tree hole mosquito \emph{Aedes aegypti} are
mainly limited by the development during the larval stage. The two
main environmental factors that have been so far identified are temperature
\citep{shar77,rued90} and food. It is expected that genetics will
play a role but comparative studies of quantitative traits are not
abundant. Studies comparing developmental rates of \emph{Ae ae.} collected
at different locations in Argentina and those of the Rockefeller strain
(a laboratory strain), indicate \citep{Grech2010} that there are
significant differences, under the same temperature and laboratory
feeding conditions, between local strains and the Rockefeller strain,
as well as statistically significant differences between local (country
wide) strains, despite being all of them identified in the same branch
of the phylogenetic tree as hybrids between \emph{Aedes aegypti aegypti}
and \emph{Aedes aegypti formosus} \citep{glor16}. For example, the
Rockefeller strain spends less time as larvae. Yet, the most substantial
difference corresponds to the total number of eggs and the daily fertility
which is five times larger for the Rockefeller strain than for the
local strains. A recent study performed in Trinidad and Tobago \citep{chad16}
reports a fertility of local females raised in the laboratory, with
local genetics, which is in line with the experiments reported in
\citep{Grech2010}.

Any model (be it mathematically explicit or not) aimed at exploring
the possible outcome of an environmental intervention, must include
the reaction norms \citep{hamilton2011population} of the local, the
released and the resulting hybrid mosquitoes. The traits that have
been identified so far are: hatching inhibition as a response to low
levels of food (the Gillett effect) \citep{gill55,gill55b,gill59,gillet77,livd87,edge92,edge93},
duration of the larvae stage \citep{sout72,rued90,fock93a,Macia2006,maci09,rome15},
mortality of preimaginal forms \citep{Macia2006,maci09,rome15} and
fertility \citep{arri04}, directly associated to the body size of
females. Furthermore, it has been observed that temperature and food
are not independent factors \citep{Padmanabha2011}.

Needless to say, the feeding behaviour of the larvae will alter the
food resources in the container. Hence, food and phenotype must be
considered as dynamical variables.

\subsection{Simple population dynamic model to evaluate effectiveness and gene
diffusion}

\subsubsection{Comments on previous models}

In Table \ref{tab:Previous-models} we display a few mathematical
models that have bearing on the issues discussed in this work. The
table classifies the models according to key features of our modelling
strategy. $Pop$ denotes the number of compartments. A large number
of  compartments allows for finer graining, since compartments differentiate
groups of individuals with different characteristics, different dynamical
responses and/or different biology. $Var$ indicates whether modelling
is performed with continuous or discrete variables. Populations are
counted with non-negative integers and are hence intrinsically discrete;
continuous variables may constitute a fair approximation to large
populations, but they are in conflict with the possibility of extinction
of one or more compartments. $Dyn$ indicates the type of dynamics.
All models but the last in the table are deterministic (either differential
equations or difference equations on discrete time). Such strategy
forces in practice the use of continuous variables. \emph{SkeeterBuster},
the last entry of the Table has an intermediate approach, being the
time-lapse of the preimaginal stages stochastic. $Eco$ describes
whether the model takes into account the interaction of the populations
with the environment. In fact, model parameters have a biological
meaning, at least metaphorically, but in general they express the
action of the environment on the populations, rather than a true interaction
where the dynamics of the population modifies the environment as well.
$Sp$ indicates whether the spatial distribution is taken into account.
Homogeneous environments can be modelled without spatial resolution,
contrasting with inhomogeneous environments. $Int$: indicates whether
introgression is considered. $GS$: denotes the genetic strategy in
the model (LG=lethal gene, Ge=Generic).
\begin{table}[h]
\begin{tabular}{|c|c|c|c|c|c|c|c|}
\hline 
Reference & $Pop$ & $Var$ & $Dyn$ & $Eco$ & $Sp$ & $Int$ & $GS$\tabularnewline
\hline 
\hline 
\citep{este05} & 5+1 & C & Det & no & no & no & Ge\tabularnewline
\hline 
\citep{phuc07} & 1+delay & C & Det & no & no & no & LG\tabularnewline
\hline 
\citep{yakob2008aedes} & 1$^{a}$+delay & C & Det & no & yes & no & Ge\tabularnewline
\hline 
\citep{li2009simple} & 2$^{a}$ & C & Det & no & no & no & Ge\tabularnewline
\hline 
\citep{white2010modelling} & 2+delay & C & Det & no & no & no & Ge\tabularnewline
\hline 
\citep{lee2013modelling,lee2013optimal} & 2 & C & Det & no & yes & no & Ge\tabularnewline
\hline 
\citep{mago09,legros2012assessing,robert2014mathematical} & $^{b}$ & D & Mix & yes & yes & no & Ge\tabularnewline
\hline 
\end{tabular}

$^{a}$ Time is discrete.

$^{b}$ Not a compartmental model.

\caption{\label{tab:Previous-models}Previous models. $Pop$: Number of populations,
$Var$: Type of variables (D=Discrete, C= Continuous), $Dyn$: Dynamics
(Deterministic, Stochastic or Mix, see text), $Eco$: Ecology, $Sp$:
indicates whether the spatial distribution is taken into account,
$Int$: Introgression, $GS$: Genetic strategy (LG=lethal gene, Ge=Generic).}
\end{table}

Concerning the central issues of this work, these models were not
conceived to address the possibility of hybridisation and its eventual
propagation in time, or the interruption of the interventions, and
are hence insufficient to discuss the questions of interest for this
study. The model in \citep{yakob2008aedes} includes a discussion
about fitness of the RIDL released males but the context is how to
compensate deficiencies in fitness with a higher release rate. The
model proposed in \citep{phuc07} is the only one addressing the LG
strategy. It has a low level of detail since it deals only with one
population. Its goal was to address questions regarding the effectiveness
of the proposed technique. FDA (Food and Drug Administration, US)
\citep{fda16} refers to this model in their assessment of the RIDL-SIT
technique. Finally, most models treat the environment as ``fixed''.
Even if biological parameters in the models are e.g., temperature
dependent, the environment is imposed on the populations without feedback,
being in this way outside the dynamical evolution. The possibility
of (local) environmental modification originated in the time-evolution
of the populations is absent.

A separate paragraph deserves the approach of \emph{SkeeterBuster}
(last entry of the table). The model produces detailed simulations
based upon Focks' model CIMSiM \citep{fock93a,fock93b}. The original
model was revised and improved to incorporate the discussion of some
genetic aspects, mainly the fixation index, $F_{ST}$, in relation
to spatial dispersion. The computation is performed on one neutral
locus (two alleles). One locus accounts only for $(00,01,11)$ classes,
meaning that wild and hybrids with no LG are considered identical.
No hybridisation can be tracked in this form. The model is highly
detailed (it tracks individual females) but its biological content
targets other questions as compared with the present work. Additionally,
the information required to run the model is difficult to obtain,
excpet for the area of Iquitos (Perú), which is the main target for
\emph{SkeeterBuster}. Iquitos has been the locus of a long term research
project in dengue and \emph{Aedes aegypti}, but even in such conditions,
parametrisation of the model is difficult: We read in the conclusions
of \citep{Ellis2011}, ``CIMSiM and DENSiM are most sensitive to
parameters for which little empirical information exists, and although
this information can be extremely valuable for setting research priorities,
it also places constraints on the use and interpretation of model
outputs.'' When compared to early versions of aedesBA (a compartmental
model that uses an intermediate level of information) no definite
advantage was observed for any of the models \citep{legr16}. The
difference between modelling philosophies has been briefly discussed
in \citep{rome13}.

Mathematical models do not produce biological data, they articulate
information to produce understanding and insight about what to expect
given the present level of information. However, knowledge at times
takes the form of knowledge of our ignorance: which information do
we need that has not yet been produced? Measurements and experiments
are made to match a preconceived form of articulation. Models call
for new experiments to produce substantially superior forms of articulation
than the irrationality in habits and intuition of (raw) empiricism.

The modelling situation presents the dilemma: should we refrain from
modelling because of the lack of information to run our most detailed
codes? Such a restriction has as consequence that models which are
too limited to address the task (e.g., \citep{phuc07}), become the
only decision-making contribution provided by the modelling community.
The best is the enemy of the good, is a popular aphorism and Italian
proverb (\emph{il meglio è l'inimico del bene}) quoted by Voltaire.
In what follows we present a viable option under the current circumstances.

\subsubsection{Lack of information and modelling decisions.\label{subsec:Lack-of-information}}

In the light of the limited or uncertain information concerning the
RIDL-SIT technique, some modelling decisions were made, namely:
\begin{itemize}
\item The lack of information regarding the reaction norms to environmental
conditions such as the availability of food have been decisive in
the modelling decision of not considering high-level models such as
\citep{rome15} which incorporate phenotype variation in terms of
environmental conditions. High-level models are sensitive to differences
found in the phenotype for eggs collected at different places \citep{rome13}.
\item The effect of the lethal gene seems to be apparent not before \emph{L4}
stage and it is strongest at pupae level (see Table 1 in \citep{phuc07}).
Hence, the relation between mortality to emergence related to unmodified
pupae is adjusted so that for individuals carrying the lethal gene
the probability of death as pupae is $\frac{19}{20}=0.95$ and the
probability of adult emergence is $\frac{1}{20}=0.05$, since these
are the only competing events in an exponential race \citep{durr01}
at pupa stage.
\item Another substantial missing information concerns the contributions
of dead larvae and pupae as a secondary source of food. Such contribution
was incorporated in earlier models \citep{fock93a} based upon results
in \citep{gilp79} and amount to a $12\%$ recovery (the product of
40\% nutritive value and 30\% conversion of food in weight) according
to these sources. Recycling of dead larvae was assumed to influence
food availability during the developmental stages (see Section \ref{subsec:Specific-details}
for a description of the food cycle).
\item Heritability of the survival trait despite the presence of the lethal
gene has been set to zero.
\item Mating effectiveness as compared with the wild strain has been rounded
up to $C=0.06$.
\item We have made no difference in the survival of adult males of different
strains (other than the food deficit effect for individuals raised
in the wild). None of the reported experiments checks lifespan of
modified males vs. lifespan of wild males in the same wild environment
under similar conditions. In our simulations, that target for a pre-established
reduction level of the wild population, the mortality of released
males will affect the release rate since a comparatively high proportion
of males must be maintained throughout the process. Eventually, larger
release rates may enter in conflict with the factory production capabilities.
The assumption made here underestimates the strain put on the factory.
\item Since there is no information regarding reaction norms of hybrids
between local strains of \emph{Ae ae.} and the Rockefeller strain,
we have initially assigned to all wild and hybrid mosquitoes the same
reaction to food deficit stress using the set of reaction norms available
for mosquitoes collected at Buenos Aires (Argentina) \citep{rome15}
corresponding, in the present model, to: pupation rate, fertility,
mortality at the preimaginal stadiums, as well as the Gillett effect
as quantitatively described in \citep{edge92} and the mortality of
adults. However, released adult mosquitoes (overwhelmingly males)
are not raised under food deficit stress and their mortality rate
is not corrected by food deficit but fixed as indicated in the previous
point.
\item In Subsection \ref{subsec:hyb} we explore to some extent the dependence
of hybridisation with respect to hybrid fitness in two ways: (a) We
test two values for the mating efficiency of heterozygous males carrying
one LG, $O1$, setting it equal to that of the wild strain or, alternatively,
to that of homozygous males (carrying two copies of the LG). (b) For
adult females, we tested doubling the mortality rate for the $O1$
hybrids. In all cases, there is no point in considering hybrids not
carrying a LG worse fitted than $O1$ mosquitoes. For hybrids not
carrying a LG (named $Rock$ in Table \ref{tab:Hybridisation-levels.-})
we considered a linear interpolation between the two extreme cases
just described, modifying at the same time the mating efficiency of
males and the mortality rate of females.
\end{itemize}
A model is not just a mere transcription of experimental findings
but a logical construction as well. In our schematic construction
we have considered the two food regimes described in \citep{rome15}
and the dependence of phenomenological traits with the logarithm of
the food concentration. A region where the life form is impossible
was considered for food concentrations below $2^{-8}$ of the optimal
concentration values obtained in the laboratory (which correspond
to $1$ in these units). The region of high mortality was considered
in the range $[2^{-8},2^{-6})$ while the favourable region spans
the interval $[2^{-6},1]$. In the most unfavourable region the only
life form possible are dormant eggs inhibited from hatching by the
Gillett effect (see Section \ref{subsec:Genetics-Environment}), all
other preimaginal forms have probability one of dying. In the transition
region, the Gillett effect diminishes from total inhibition to no
inhibition and the mortality moves from probability one to the values
measured in laboratory experiments \citep{rome15}. The fertility
is considered to be proportional to the mass excess to the minimal
form of the adult found at the $2^{-8}$ food concentration where
it takes a zero value and grows to reach the measured value for \emph{Ae
ae.} in the region of optimal laboratory conditions \citep{Grech2010}.
Laboratory results indicate that the increase in size as measured
by the wing length is linear with the logarithm of the food concentration
\citep{rome15b} all along the considered region. Using the raw data
of the experiment reported in \citep{rome15,rome15b} we determined
that the event rate as larvae (considering both death and pupation)
increased linearly (with the logarithm of food concentration) in the
favourable region and was approximately constant in the transition
region.

\subsubsection{Stochastic Population Dynamics\label{stopopdyn}}

This approach to population studies assumes that the dynamics is adequately
described by classifying the population into \emph{compartments} $X_{j}$,
$j=1,\cdots,N$ (described by nonnegative integer numbers), each compartment
counting the number of individuals of a given type as a function of
time. For example, mosquito subpopulations in the present study are
represented by the class $D$ with elements $\lbrace$egg, larvae,
pupae, adult female, adult male$\rbrace$. Each element in $D$ will
be further subdivided with respect to its genetic type according to
$G$, with elements $\lbrace$wild, hybrid, modified heterozygous,
modified homozygous$\rbrace$. Thus, the index $j$ can be thought
of as listing the elements of the external product $D\times G$. Changes
in time occur stochastically and are identified by \emph{events},
labelled by the index $\alpha$, where $\alpha=1,\cdots,E$. The event
$\alpha$ produces an instantaneous change in population compartment
$j$ given by the integer $\delta_{j}^{\alpha}$. Note that events
where an individual undergoes a transition from one compartment to
another (e.g., pupation where a larva becomes pupa) have exactly two
nonzero indices $\delta_{j}^{\alpha}$: For $j=l$ (larva) $\delta_{l}^{\alpha}=-1$,
while for $j=p$ (pupa) $\delta_{p}^{\alpha}=+1$. Also, death events
have $\delta=-1$ for the corresponding compartment. Birth events
(oviposition) may have $\delta>1$(see Subsection \ref{subsec:Specific-details}).
The time-evolution of the population, given an initial condition (here
labelled as $t=0)$ can be summarised as:
\[
X_{j}(t)=X_{j}(0)+\sum_{\alpha=1}^{E}\delta_{j}^{\alpha}n_{\alpha}(t)
\]
where $n_{\alpha}(t)$ is the number of events of type $\alpha$ that
occurred up to time $t$ (a stochastic variable). This approach is
called a Markov Jump Process (\emph{Markov} since the only necessary
information to compute the next change is the knowledge of the present
state; \emph{Jump} since the changes that take place are described
by the integer $\delta_{j}^{\alpha}$ indicating a change in the number
of individuals in compartment $j$). In the mathematical literature
attention to these processes starts with Kolmogorov’s foundational
work \citep{Kolmogoroff1931} and its further elaboration by Feller
\citep{Feller1940}. A substantial effort to relate the stochastic
description to deterministic equations was performed by Kurtz \citep{ethi86,kurt70,kurt78}.

Traditionally, attention was focused on the dynamics in population
space, rather than event space. In the case of probability rates that
are linear in the subpopulations both approaches are quite interchangeable\citep{sola14}.
However, event space is more appealing for a general understanding.
For example, when the dynamics is affected by introducing a treatment
or strategy regarded as an event (think e.g., of vaccination in an
epidemic disease) it will be important for policy reasons (logistics,
costs, etc.) to count the number of occurrences of that event, rather
than attempting to sort it out by reversing the information contained
in the population set $X=\left\{ X_{j}(t),\,j=1,\cdots,N\right\} $.

The Markov assumption leads to defining probabilities per unit time
$W_{\alpha}$ (in the sequel called \emph{rates} or \emph{probability
rates}) for each event. In this work it will suffice to consider that
these probability rates depend only on the population $X$ at each
given time (other dependencies will be handled along the way). In
a small time-interval $h$ the probability of occurrence/non occurrence
of event $\alpha$ is the pair $\left(W_{\alpha}(X)h+o(h),1-W_{\alpha}(X)h+o(h)\right)$,
where $o(h)$ is a quantity that goes to zero with $h$ faster than
linearly.

The process is further modelled implementing the Feller-Kendall algorithm
\citep{kend49,kend50} that can be summarised in the following steps
\citep{sola14}:
\begin{enumerate}
\item Starting at $t=0$ or immediately after an event has occurred, the
waiting time to the next event is exponentially distributed, with
rate $R=\sum_{\alpha}W_{\alpha}(X)$ equal to the sum of all event
probability rates, where $X$ refers to the population values after
the most recent event has occurred (or to $X(0)$). The time $\tau$
to the next event is simulated by picking an exponential random deviate
with rate $R$.
\item Given that an event has occurred at time $\tau$, the probability
of it being event $\alpha$ is
\[
P(\alpha/\mathrm{event}\,\mathrm{at}\,\tau)=\frac{W_{\alpha}(X)}{\sum_{\alpha}W_{\alpha}(X)}=\frac{W_{\alpha}(X)}{R}.
\]
 Choosing an arbitrary order for the events, the event that took place
at $\tau$ is simulated by picking a uniformly distributed random
number $Y$ in $[0,R]$ and checking to which event in the order it
corresponds. For example, if $\sum_{\alpha=1}^{K}W_{\alpha}(X)<Y<\sum_{\alpha=1}^{K+1}W_{\alpha}(X)$,
we assign the occurrence to event $K+1$.
\item Populations are updated according to $\delta_{j}^{\alpha}$ for the
assigned event and a new cycle 1-3 is computed until the final simulation
time is reached.
\end{enumerate}

\subsubsection{Schematic description of the model}

The setup in consideration is that within a limited natural population
of \emph{Ae ae}., male adult mosquitoes are introduced (with an error
of 1 female every 4300 releases \citep{carv15}). These mosquitoes
differ in two ways from the wild population. Firstly, they have two
copies of a dominant LG (not present in the wild population) that
induces massive death of the offspring before adult stage. The released
mosquitoes have been previously raised in the laboratory, from the
Rockefeller strain of \emph{Ae ae.}, a laboratory-adapted strain.
These mosquitoes are genetically different from the original wild
population. If the totality of the offspring does not die before reproduction,
there will be some degree of population mixing and genetically modified
individuals (relative to the original wild population) could arise.

Hence, to describe this problem a model has to be designed covering
three goals: 
\begin{enumerate}
\item Describe the wild population with sufficient accuracy taking care
of environmental constraints. 
\item Take care of the evolution of the LG in the offspring along generations.
A released homozygous mosquito mating with a wild female (where the
corresponding gene is not lethal) will yield heterozygous offspring
and any eventual subsequent offspring will propagate this special
gene according to the mendelian rules.
\item Take care of the mixing between the natural population and the released,
genetically different, mosquitoes (mixing of other genes apart from
the lethal one).
\end{enumerate}
Within this approach, the specification of the model requires (a)
to identify the populations (compartments) participating in the dynamics
and (b) to specify the events that define the dynamics, their reach,
biological content and environmental conditions. 

As mentioned in the previous Subsection, the subpopulations involved
are chosen according to the developmental stages (egg, larva, pupa,
male adult and female adult). In the absence of released mosquitoes
the events governing the population dynamics are egg hatching and
mortality, larval pupation and mortality, pupa emergence and mortality,
fecundation (mating) and oviposition and finally adult mortality.
These events take care of modelling goal i.

All compartments are further classified according to their genetical
content. There is a compartment for the wild lineage and three for
the mixed lineage where $0$, $1$ or $2$ copies of the LG may be
present. This population space allows for dealing with modelling goal
ii.

Finally, each subpopulation has attached a number in $[0,1]$ representing
the average percentage of foreign genetic material. Released mosquitoes
contribute with $1$, wild mosquitoes with $0$ and the mixing propagates
according to the law of independent assortment of alleles \citep{hamilton2011population}.
This takes care of modelling goal iii.

\begin{figure}[h]
\centerline{\includegraphics{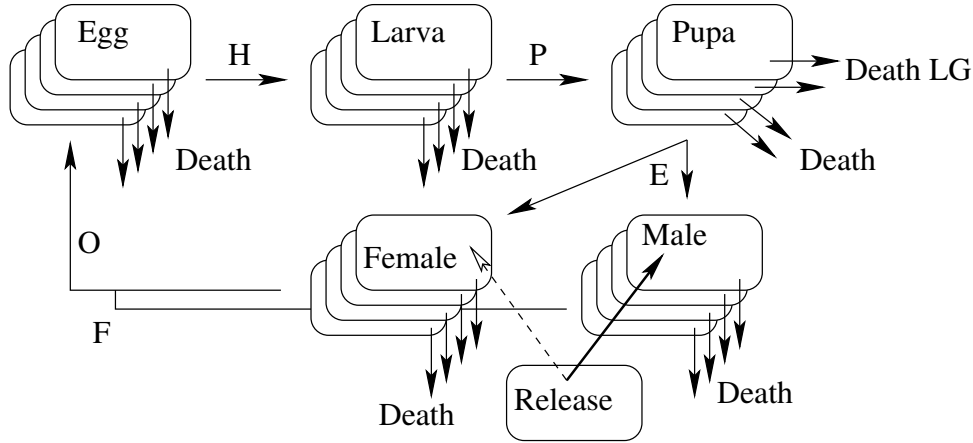}}

\caption{Compartments and events. The four copies of each compartmental stage
correspond to the cases wild, 0, 1, or 2 copies of the LG. H: hatching,
P: pupation, E: emergence, F: fecundation (mating), O: oviposition,
Death: Death rate (specific for each stage), Death LG: modified death
rate at the pupa stage due to the presence of the LG.\label{fig:Compartments}}
\end{figure}

A detailed scheme of compartments and relating events is shown in
Figure \ref{fig:Compartments}.

\subsubsection{Food dynamics}

Most rates, especially larval pupation and mortality rates, depend
on the availability of food. Following the model in \citep[Romeo Aznar et al.][]{rome13}
we assume that the totality of oviposition sites can optimally host
$l_{opt}$ larvae (grown to their potential size), that may feed unrestrictedly.
At the same time, food in the sites is produced by (environmental)
bacterial activity, it degrades at a certain rate $u$ (also depending
on the environment, in this work $u=\frac{1}{days}$) and it is consumed
at a pace depending on the amount of larvae. Dead larvae and pupae
are to some extent recycled as available food, as mentioned in Section
\ref{subsec:Lack-of-information}. Let $\frac{P_{f}}{C_{f}}$ denote
the ratio of produced to consumed food. Initially, we set its value
to the deterministic equilibrium value, an environment-dependent estimate.
We denote the leftover food (relative to consumed food) as $L_{f}$.
In conditions of food scarcity, $L_{f}=0$. The dead larvae or pupae
recycled as food is denoted as $X_{l}$. This last parameter is initially
zero and it is updated at every larval or pupal mortality event (see
below). The ratio $\frac{P_{f}}{C_{f}}$ indicates the environmental
scarcity conditions: if lesser than one, there is no food available
for the larvae to reach the potential weight at pupation (this ratio
enters in several steps of the modelling procedure). The food-cycle
for each simulation time-step $\tau$, where the initial total larva
population is $L$ and the ideal estimated capacity of the oviposition
sites is $l_{opt}$, proceeds as follows: 
\begin{eqnarray*}
L_{f} & = & max(\frac{P_{f}}{C_{f}}-1,0)\\
\frac{P_{f}}{C_{f}} & = & L_{f}\,exp(-u\tau)+\frac{l_{opt}+X_{l}}{L}\\
X_{l} & = & X_{l}exp(-u\tau)
\end{eqnarray*}
At each time-step, the values of $L$ and $X_{l}$ are updated and
all above quantities recalculated. The influence of the food in mosquito
dynamics is described by larval event rate $LE(x)$, mortality probability
as larvae and pupae $ML(x$) , hatching inhibition $GL(x)$ and fertility
rate $FT(x)$
\begin{eqnarray*}
LE(x) & = & \max(0.025\,,\,0.25269+0.031974\,x)\\
ML(x) & = & \begin{cases}
1 & x<-8\\
\max(0.033\,,\,0.033-0.4835\,(x+6.)) & x\ge-8
\end{cases}\\
GL(x) & = & \begin{cases}
1 & x\ge-6\\
(1.+0.5\,(x+6.)) & -8\le x<-6\\
0 & x<-8
\end{cases}\\
FT(x) & = & \max(0.\,,\,0.127742\,(x+8.))
\end{eqnarray*}
where $x=\log_{2}(\frac{Cf}{Pf})$. The fertility rate changes the
fertility of the emerging females, and then the average fertility
of females (see below). The values for the reaction norms correspond
to the values reported in \citep{rome15}. In addition, the average
lifetime as adult is corrected as $m_{a}=\frac{m_{a0}}{1-ML(x)}$
obtaining in this form a reasonable correspondence between times measured
at the laboratory (see for example \citep{chad16}) and in field experiments
(for example \citep{sout72}). Released male mosquitoes, that had
no food deficit while raised, were set to have average lifetime $m_{a0}$
(see values below).

\subsubsection{Specific details\label{subsec:Specific-details}}

The theoretical basis of the model when it comes to the population
dynamics of \emph{Ae ae.} refers to \citep{rome13,oter06,oter08}.
The problem is \emph{quasi-linear} \citep{sola14} in the populations,
meaning that (a) all events that decrease a subpopulation produce
a decrease in one unit in only one population compartment and (b)
all non constant rates can be written in the form $W_{\alpha}=m_{\alpha}^{i}X_{i}$,
for the relevant population $X_{i}$. However, the coefficients $m_{\alpha}^{i}$
need not be all constant. 

The events \emph{hatching} and \emph{pupation} are assumed to have
the same coefficient for all four compartments of the participating
subpopulation. All these events pick one individual from the original
subpopulation and passes it to the corresponding subpopulation of
the next stage. Pupa emergence is less likely for the compartments
carrying $1$ or $2$ copies of the LG (see below for the explicit
values). Emergence events are randomly assigned to yield $50\%$ males
and females. Each \emph{mortality} event simply decreases the corresponding
subpopulation in one unit. Except in the case of pupa mortality, where
the effect of the LG also distinguishes two cases, the mortality coefficient
is taken to be the same along each stage. The pupa mortality is much
higher for the compartments having $1$ or $2$ copies of the LG (see
Appendix I for details of the estimates). As indicated in Section
\ref{subsec:Lack-of-information}, adult mortality was assumed to
be equal for all strains. This modelling decision may overestimate
the size of non wild subpopulations, an effect that may be mitigated
by larger releases. 

The \emph{release} was acted weekly during the treatment period, following
the implementations of this technique discussed above, that only report
weekly releases. The release goals of \citep{harr12} were a relation
of $10:1$ in adult males (later increased to $25:1$) and $50\%$
fluorescence in the larvae from ovitrap-collected eggs. The initial
size of the release is therefore set to be $11$ times larger than
the adult male population in our computations. Later release events
are increased or reduced adaptively every week, in order to adjust
the weekly proportion of eggs with LG against a target value. There
is however a maximum release size corresponding to the maximal capacity
of the production plant. In this work it is set to $100000$ individuals
per release. Release is intended to concern only modified males, but
a few females are released as well, since the accuracy of the separation
technique is limited (one female every 4300 releases \citep{carv15}).

The model simulates a \emph{closed environment}, i.e., no emigration/immigration
of adult individuals to and from adjacent areas is taken into account.
Clearly, such effects could only delay or reduce any effects of the
release since immigrants, at least initially, would be wild individuals
while emigrants could be any among released, wild and mixed individuals. 

In this implementation, fecundation (mating) is not treated as a separate
event. Instead, we assume that given that the adult male population
is nonzero, all females are fecundated immediately after emergence.
Thus, each emerged female has two associated indices, corresponding
to the adult female compartment (inherited from the pupa) and the
fecundating male compartment. The latter is chosen randomly among
the four male subpopulations, weighing the size of the subpopulation
with two copies of the LG with a factor $0.06$ as described in the
previous Sections. We assume hence that this reduced effectiveness
is coupled to a high proportion of non-wild genes and to the homozygous
property.

The index pair given by fecundation is used to propagate to the offspring
both the LG and the percentage of introduced genes from the released
population (according to the independent assortment of alleles \citep{hamilton2011population},
i.e., with the average of parental percentages). The propagation takes
place in the oviposition event that adds individuals to the egg population
(distributed among the four egg compartments according to mendelian
rules) without decreasing any other subpopulation. The amount of eggs
laid by each female at each oviposition depends on temperature and
on the size of the female, which in turn depends on the environmental
food availability. In average, wild mosquitoes collected in Argentina
and reared in an environment with optimal food availability ($\frac{P_{f}}{C_{f}}=1$)
at $T=26\,C$, lay $2$ eggs per day \citep{Grech2010}. Hence, the
number of eggs per oviposition is $\delta_{egg}=\frac{2}{m_{ovi}(26)\,FT(1)}FT(\frac{P_{f}}{C_{f}})\equiv c_{lay}FT(\frac{P_{f}}{C_{f}})$.
Once a female has emerged, its size is fixed as well as its egg-laying
capacity, both by the value of $\frac{P_{f}}{C_{f}}$ at emergence
(which varies among events according to food dynamics). Thus, at each
emergence event, $\delta_{egg}$ is updated as follows:
\begin{equation}
\delta_{egg}:=\frac{\delta_{egg}F+c_{lay}FT(\frac{P_{f}}{C_{f}})}{F+1},\label{eq:eggs}
\end{equation}
where $F$ is the total number of adult females existing previous
to the actual emergence. This is a sort of ``moving average''.

Dead larvae and pupae are partially recycled by the environment within
the hatching sites. For each death event we add to $X_{l}$ the amount
$\Delta l=xl\frac{FT(\frac{P_{f}}{C_{f}})}{FT(1)}.$ The quotient
estimates the reduced size of each dead larva or pupae relative to
ideal food conditions. The values $xl=0$, $0.12$ (as suggested by
results in \citep{fock93a} and \citep{gilp79}, see Section \ref{subsec:Lack-of-information})
and $0.5$ were tested.

Most rates are known to be temperature dependent \citep{rued90,oter06,Padmanabha2011}.
In this work we fixed temperature at the value $T=26\,C$, corresponding
to a year-round stable tropical environment. The coefficients for
the different rates (in units of $(days)^{-1}$) are displayed in
Table \ref{tabla-rates}, along with the references where the model
structure and rates have been discussed:

\begin{table}[h]
\setlength\tabcolsep{2pt}\begin{threeparttable}%
\begin{tabular}{|c|c|}
\hline 
Event & Rate\tabularnewline
\hline 
\hline 
Hatching$^{a,b}$ & $m_{e\to l}=\left(-0.0167105+0.03866\,\exp(\frac{T+7.26}{16.47906})\right)GL(\frac{P_{f}}{C_{f}})$\tabularnewline
\hline 
Egg mortality$^{a,c}$ &  $m_{e}=0.01$\tabularnewline
\hline 
Pupation$^{b}$ &  $m_{l\to p}=LE(\frac{P_{f}}{C_{f}})\,\left(1-ML(\frac{P_{f}}{C_{f}})\right)$\tabularnewline
\hline 
Larval mortality$^{b}$ & $m_{l}=LE(\frac{P_{f}}{C_{f}})\,ML(\frac{P_{f}}{C_{f}})$\tabularnewline
\hline 
Pupa emergence no LG$^{b}$ & $m_{p\to a}=0.5787\,\left(1-ML(\frac{P_{f}}{C_{f}})\right)$\tabularnewline
\hline 
Pupa mortality no LG$^{b}$ & $m_{p}=0.5787\,ML(\frac{P_{f}}{C_{f}})$\tabularnewline
\hline 
Pupa emergence with LG  & $m_{pl\to a}=\frac{1}{20}\left(m_{p}+m_{p\to a}\right)=\frac{1}{20}\,0.5787$\tabularnewline
\hline 
Pupa mortality with LG & $m_{pl}=\frac{19}{20}\left(m_{p}+m_{p\to a}\right)=\frac{19}{20}\,0.5787$\tabularnewline
\hline 
Adult mortality$^{b}$ &  $m_{a}=\frac{0.04}{(1-ML(\frac{P_{f}}{C_{f}}))}$\tabularnewline
\hline 
Adult mortality (2 LG)$^{d}$ & $m_{a}=0.04$\tabularnewline
\hline 
Oviposition$^{b}$ &  $m_{ovi}=0.03154\,exp(\frac{T-4.7511}{10.580})$\tabularnewline
\hline 
\end{tabular}

\begin{tablenotes}
\item[a] See \citep{oter06}. $^b$ See \citep{rome13,rome15b}. $^c$ See \citep{trpi72}. $^d$ See \citep{fda16}.
\end{tablenotes}
\end{threeparttable}\caption{Events and event rates. (LG stands for Lethal Gene) \label{tabla-rates}}
\end{table}

\subsubsection{Hybridisation degree}

The inheritance of genotype is acted at fecundation. On one hand,
the different subpopulations are classified in four compartments,
depending on they being ``purely wild'' or mixed (e.g., a wild female
mating a released male), with $0,\,1$ or $2$ copies of the LG (here
labelled $W,\,M_{0},\,M_{1},\,M_{2}$). When the treatment starts,
the first generation offspring of the released males with wild females
will invariably land on compartment $M_{1}$ but later generations
may mix further, since pre-adult mortality of the mixed offspring
is not total. In any case, the offspring falls always in one of the
four compartments.

Regarding the rest of the genetic material, the average degree of
mixing $R$ of each compartment is computed in several steps. This
value changes for each individual at fecundation/oviposition. Further,
when new individuals enter a compartment, the compartmental average
for $R$ changes accordingly and this modification propagates along
the life cycle. Wild population has initial mixing value $0$, while
released mosquitoes have value $1$, corresponding to $100\%$ Rockefeller-strain
genetic material. Following the law of independent assortment of alleles,
offspring has a mixing equal to the average of the parental values,
$R_{o}=\frac{1}{2}\left(R_{m}+R_{f}\right)$. Laid eggs (eq. \ref{eq:eggs}),
however, distribute among compartments according to mendelian laws.
Finally, the average mixing of the affected compartments is consequently
moved as, 
\[
R_{n}:=\frac{R_{n}X(n)+R_{o}E}{X(n)+E},
\]
where $E$ is the amount of eggs (with inherited mixing $R_{0}$)
belonging to compartment $n$ and $X_{n}$is the preexistent number
of eggs in that compartment (with previous average mixing $R_{n}$).

\section{Results\label{sec:Results}}

The code used for the simulations is deposited on GitHub. See \url{https://github.com/MNatiello/gene-diffusion-mosquito}

Model simulations were performed using the following scheme: 
\begin{itemize}
\item The initial wild populations and environmental food conditions were
set to the deterministic equilibrium value (see Appendix II for a
description).
\item For a transient time $t_{T}=156$ weeks (about $3$ years), the wild
population was allowed to proceed without interference. The population
data was saved to start the next simulation.
\item After the previous transient, intervention was started, for a duration
$t_{I}=52$ or $260$ weeks for the results of Subsection \ref{subsec:Summary1}
and $t_{I}=116$ weeks for those in Subsection \ref{subsec:hyb}.
\item After finishing the intervention, the system was simulated for another
$156$ weeks.
\item The exhibited results are the average of $10$ runs of the previous
scheme.
\item The short treatment was targeted to a weekly fluorescence ratio of
$0.5$ and the longer ones to ratios of $0.5$ and $0.65$ between
eggs hatching to fluorescent vs total number of hatched eggs (per
week).
\item The recycle coefficient $xl$ at equilibrium was set to $0.12$ for
the short treatment while for the long treatment we illustrate the
dynamics with $xl=0,\,0.12$ and $0.5$.
\item There is no allowance for differentiated hybrid adult fitness (other
than in Subsection \ref{subsec:hyb}).
\end{itemize}

\subsection{Summary of results\label{subsec:Summary1}}

We illustrate the simulations results with a few graphs indicating
the time-evolution of the displayed properties. The abscissa of all
graphs has units of days, counted from the beginning of the simulation.
The transient, treatment and final portions are evident.

\begin{figure}[h]
\centerline{\includegraphics[width=7cm]{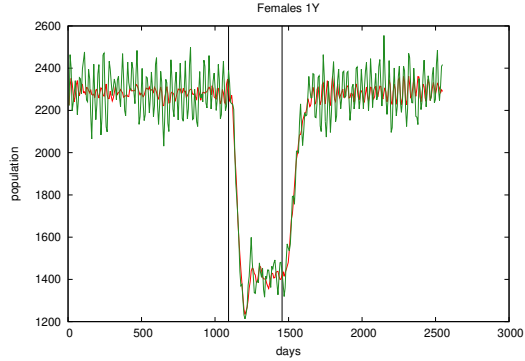}}\caption{Females with no LG, 1-year treatment, 50\% eggs with LG. Vertical
lines indicate beginning and end of treatment. Green: single simulation
line, Red: average of 10 simulation runs\label{fig:Females1}.}
\end{figure}

In Figure \ref{fig:Females1} we display the time-evolution of the
number of adult females carrying no LG for the short treatment (1
year), where the ratio of eggs with LG to total number of eggs each
week is targeted to $50\%$. The recycling coefficient of dead larvae
and pupae is set to $0.12$. The drop in population size during the
intervention is evident. Initially, the drop is more intense since
the adjusting mechanism to target fluorescent eggs is slow (see below
for a discussion). For comparison, we plot one single simulation run
superimposed to the average of 10 simulation runs.

\begin{figure}[h]
\centerline{\includegraphics[width=7.5cm]{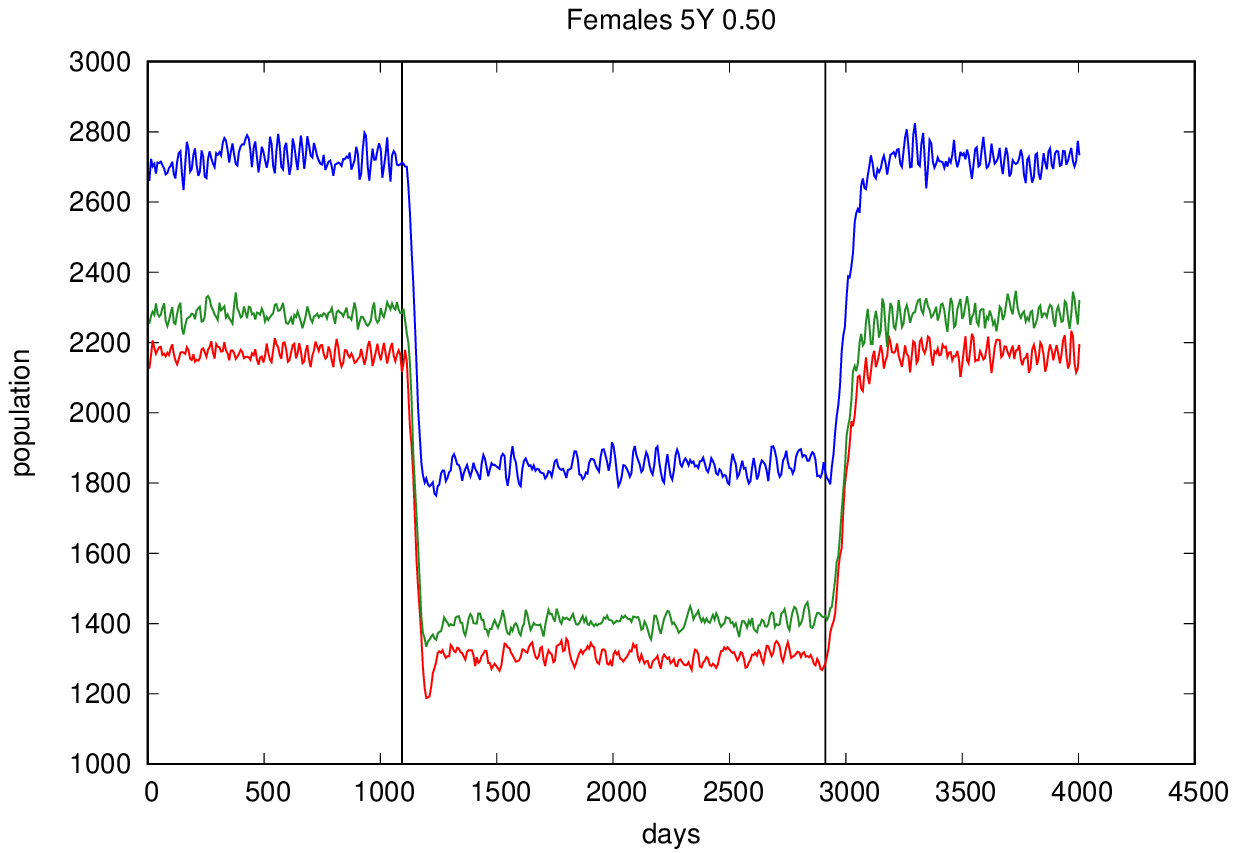}\includegraphics[width=7.5cm]{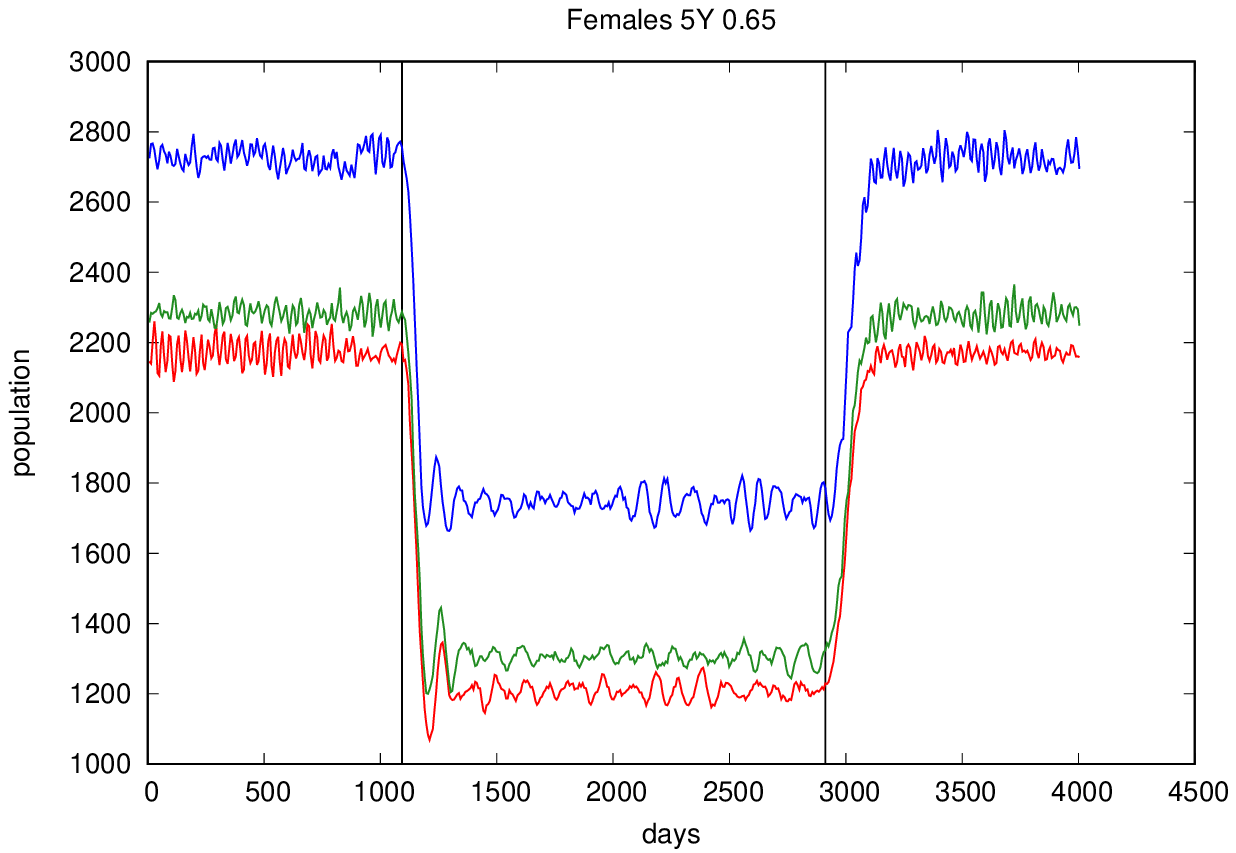}}\caption{Females with no LG, 5-year treatment, $50\%$ (left) or $65\%$ (right)
eggs with LG. Curves correspond to values $0$ (red), $0.12$ (green)
or $0.5$ (blue) for the recycling coefficient of dead larvae and
pupae. Vertical lines indicate beginning and end of treatment\label{fig:Females5}. }
\end{figure}

Figure \ref{fig:Females5} displays the time-evolution of the number
of adult females carrying no LG for the long treatment (5 years),
where the ratio of eggs with LG to total number of eggs each week
is targeted to $50\%$ or $65\%$. Same considerations as in the previous
picture hold. The three curves correspond to different values of the
recycling coefficient of dead larvae and pupae (see caption).

The target goal we attempt to simulate follows \citep{harr12} and
aims to obtain a $50
$ or $65\%$ proportion of laid eggs every week carrying one or two
copies of the LG. The idea of monitoring eggs in order to regulate
the treatment may be simple, but the control procedure is quite involved.
On one hand, eggs have to be hatched in order to observe the larval
fluorescence, meaning that the information about the system in actual
implementations is obtained with certain delay (of the order of days).
On the other hand, the control is acted by modifying the release,
i.e., mainly adult males. In the model, it takes two (target $50\%$)
or three (target $65\%$) weeks before the release of adults propagates
to a modified proportion of eggs of about half the desired target.

\begin{figure}[h]
\centerline{\includegraphics[width=7.5cm]{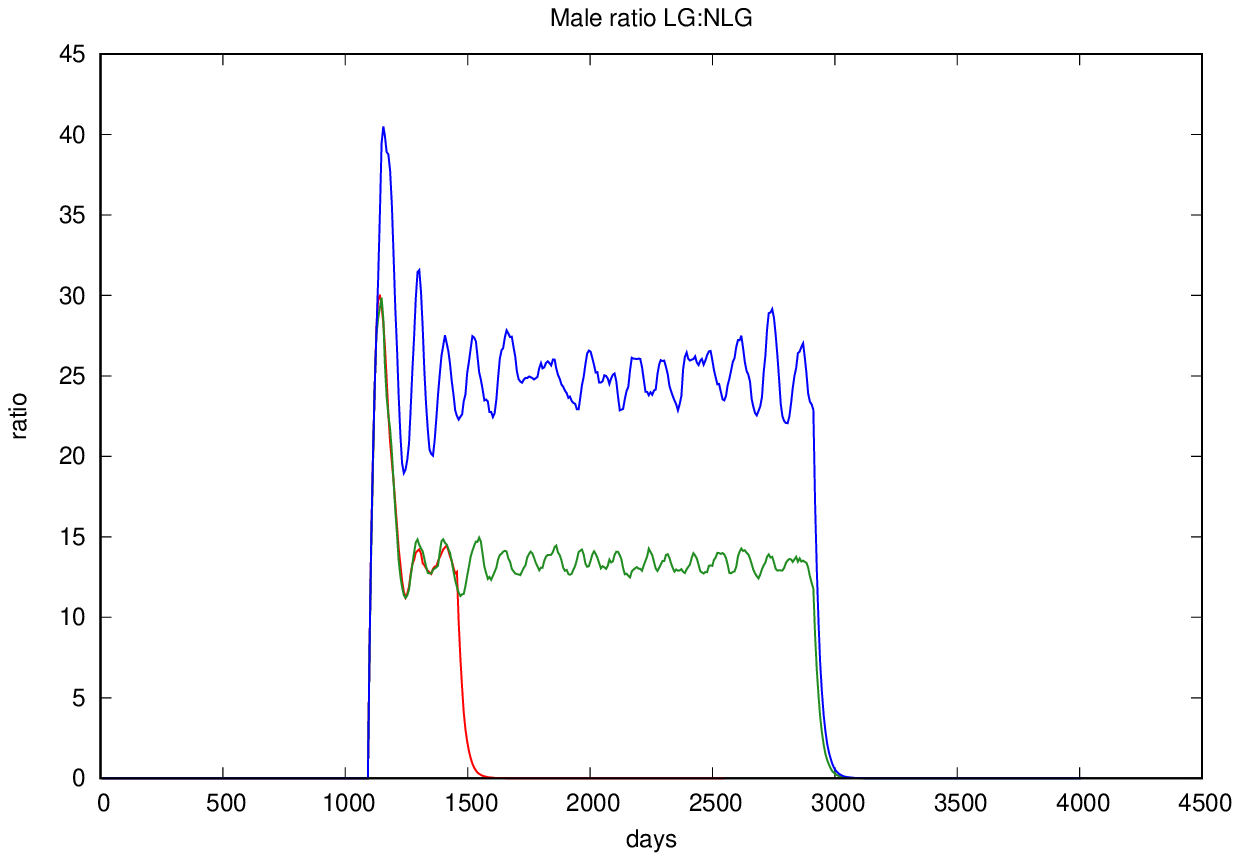}\includegraphics[width=7.5cm]{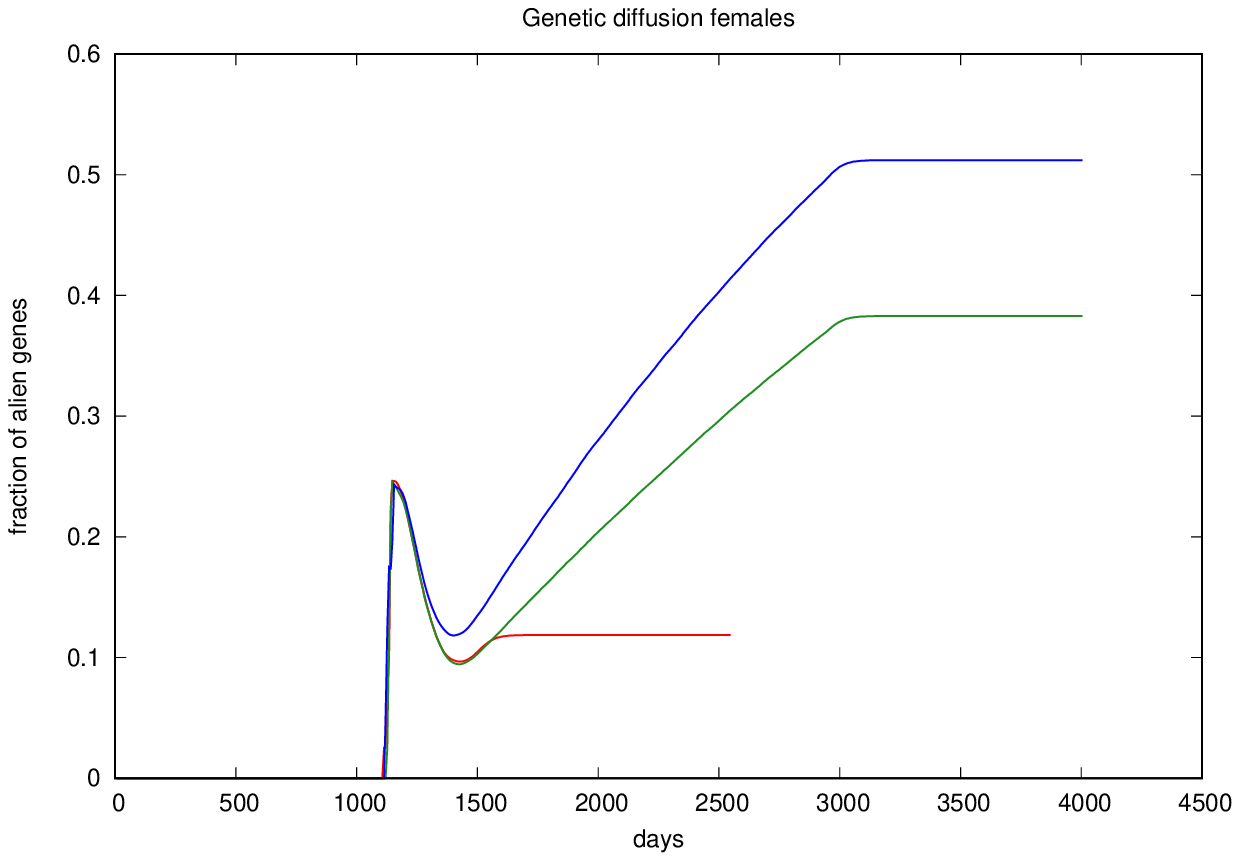}}\caption{Adult male ratio with vs without LG (left) and genetic diffusion,
i.e., the proportion of alien genes in the female population (right).
Red: 1 year, target $0.5$. Green: 5 years, target $0.5.$Blue: 5
years, target 0.65.\label{fig:graph3}}
\end{figure}

The simulations display a reduction of adult females (and all other
subpopulations except adult males) during the treatment period. This
reduction could be said to be environment dependent, since it varies
with the ability of the oviposition/breeding sites to recycle dead
larvae and pupae into more food. There is an initial drop passing
the target since the control procedure is indirect (eggs are monitored
but the only modification to the system acts through the introduction
of adult males). The other target in \citep{harr12} is satisfied
in excess, as seen in Figure \ref{fig:graph3}, left. The adult male
ratio is permanently above $10:1$ during the treatment, while in
the most demanding treatments it can be as high as $40:1$. 

After finishing the treatment, the system returns to a situation that
is comparable with the initial equilibrium condition, the main difference
being displayed in Figure \ref{fig:graph3}, right. This figure displays
the time-evolution of the proportion of alien genes in the population,
a measure for genetic diffusion. Only the recycling coefficient $0.12$
is plotted, since the differences with the other situations are minor.
Three of the four female compartments die out, remaining only the
hybrid compartment with no LG. The new equilibrium condition, although
being quantitatively similar to the one before treatment, bears now
different mosquitoes, the difference being larger for the longer and
more demanding treatments.

 The fact that the system returns to its previous equilibrium value
can be understood in different levels. From the simulation viewpoint,
the system was assumed to be in a stable equilibrium before treatment.
Despite the more or less drastic influence of the treatment, it eventually
returns to this equilibrium when left alone. However, simulations
with large weekly releases (about eight times larger than the maintenance
release for the $0.5$ target) under a sufficiently long time (more
than one year) shift the final equilibrium to a smaller permanent
population with $100\%$ Rockefeller hybridisation homozygous in the
LG. In Appendix D we comment on the effect of constant weekly release
of different sizes.

From the modelling viewpoint, we pointed out before that there exists
no knowledge about the adaptability of the modified hybrid mosquito
populations to the actual environment. This possibility is consequently
not considered in the model. Similarly, it is not known whether mosquitoes
with $50\%$ of the genetic content of the Rockefeller strain differ
from the wild strain in their behaviour and their efficiency as vectors
of viral diseases.

\subsection{Dependence of the degree of hybridisation with hybrid fitness.\label{subsec:hyb}}

The target value was adjusted to obtain a control (reduction) of the
viable-egg population close to 70\%, resulting in a value of 0.65
for our targeting strategy. In Table \ref{tab:Hybridisation-levels.-}
we show the achieved population control, the hybridisation level,
$R$, and its range (given as the extreme values of $R$ over 10 runs
of the simulation) for different control targets and recycling level
of food, while using different hypotheses regarding the fitness of
the hybrid adult population. $O1$ are hybrids with one LG and $Rock$
is the hybrid with no LG. At the end of the runs all the population
was of $Rock$ type. The intervention were chosen to be 116 weeks
long in accordance with the intervention at Juazeiro \citep{garz17}. 

\begin{table}[h]
\begin{tabular}{|c|c|c|c|c|c|c|c|c|}
\hline 
\multicolumn{2}{|c|}{$O1$} & \multicolumn{2}{c|}{$Rock$} & \multirow{2}{*}{Target} & \multirow{2}{*}{$xl$} & \multirow{2}{*}{Control} & \multirow{2}{*}{$R$} & \multirow{2}{*}{$R$-range}\tabularnewline
\cline{1-4} \cline{2-4} \cline{3-4} \cline{4-4} 
MME & FIM & MME & FIM &  &  &  &  & \tabularnewline
\hline 
\hline 
1 & 1 & 1 & 1 & 0.65 & 0 & 0.68 & 0.29 & 0.276--0.297\tabularnewline
\hline 
1 & 1 & 1 & 1 & 0.65 & 0.12 & 0.68 & 0.29 & 0.283-0.307\tabularnewline
\hline 
1 & 1 & 1 & 1 & 0.65 & 0.50 & 0.62 & 0.29 & 0.276--0.296\tabularnewline
\hline 
0.06 & 2 & 1 & 1 & 0.6 & 0 & 0.64 & 0.093 & 0.085--0.098\tabularnewline
\hline 
0.06 & 2 & 1 & 1 & 0.65 & 0 & 0.69 & 0.107 & 0.097--0.115\tabularnewline
\hline 
0.06 & 2 & 1 & 1 & 0.7 & 0 & 0.76 & 0.124 & 0.114--0.123\tabularnewline
\hline 
0.06 & 2 & 1 & 1 & 0.65 & 0.50 & 0.65 & 0.107 & 0.093--0.115\tabularnewline
\hline 
0.06 & 2 & 1 & 1 & 0.65 & 0.12 & 0.68 & 0.107 & 0.101--0.112\tabularnewline
\hline 
0.06 & 2 & 0.812 & 1.2 & 0.65 & 0.12 & 0.68 & 0.104 & 0.093-0.116\tabularnewline
\hline 
0.06 & 2 & 0.624 & 1.4 & 0.65 & 0.12 & 0.68 & 0.087 & 0.081-0.094\tabularnewline
\hline 
0.06 & 2 & 0.436 & 1.6 & 0.65 & 0.12 & 0.68 & 0.051 & 0.028--0.063\tabularnewline
\hline 
0.06 & 2 & 0.248 & 1.8 & 0.65 & 0.12 & 0.68 & {*}$^{1}$ & {*}$^{1}$\tabularnewline
\hline 
0.06 & 2 & 0.060 & 2 & 0.65 & 0.12 & 0.68 & {*}$^{2}$ & {*}$^{2}$\tabularnewline
\hline 
\end{tabular}

$^{1}$: The proportion of hybrids at the end of the simulation is
$0.45\%$ ($Rock$ only). $R$-values below roundoff limits. $R$-value
at the end of the intervention was $0.080$.

$^{2}$: $Rock$ hybrids become extinct towards the end of the simulations
(almost three years after the end of the intervention). $R$-value
at the end of the intervention was $0.016$.

\caption{\label{tab:Hybridisation-levels.-}Hybridisation levels. $O1$ mosquitoes
present one copy of LG and $Rock$ are hybrids with no deadly gen.
Simulations: 156 weeks before treatment, 116 of treatment, 156 weeks
after treatment. MME= ''Male Mating Efficacy'' FIM= ''Female Increased
Mortality''. $xl$ is the proportion of dead larvae, etc. recycled
into food (see text), Target (Control) indicates the aimed (actual)
proportion of genetically modified eggs, $R$ is the hybridisation
degree and $R$-range is given as the extreme values over 10 runs
of the simulation.}
\end{table}

\subsection{Report of simulated scenarios: Questions and answers.\label{subsec:Report}}

We summarise our findings with a set of questions and answers. The
first two questions are those present in the Introduction and originally
triggering this study.
\begin{lyxlist}{00.00.0000}
\item [{A.}] Which lasting modifications of the environment will be produced
by these tests? \\
The main modification is the hybridisation of the released mosquitoes
with the local populations as a result of genetic diffusion. The degree
of hybridisation depends both on the duration of the treatment and
its intensity (target), as well as on the (environmental) differences
in food dynamics.
\item [{B.}] Will the population of vectors previous to the environmental
intervention be re-established?\\
The recovery time of populations can be counted in months. There is
no difference in population sizes before and after the treatment for
the targeted programmes.
\item [{C.}] What is the expected ratio between released males and local
males necessary to achieve different levels of control? \\
The ratio ranges from $12:1$ to $40:1$ to sustain a target of $0.5$
to $0.65$ of the eggs having the modified genetics. Ratios larger
than $100:1$ are needed to detect sizeable effects at the beginning
of the intervention. This can be compared with the reports in \citep{harr12}
where the test started setting a ratio of $10:1$ in a comparatively
large area but it soon became clear that a smaller (or much smaller)
area and a higher ratio ($25:1$) were necessary in order to observe
some effects. Recall that the simulations assume a closed environment.
Field trials are likely to require higher ratios because of the possible
immigration of wild adults into the intervention region (see point
F.) and the emigration of released mosquitoes.
\item [{D.}] How are the results affected by the limitations of a hypothetical
mosquito factory? \\
In the simulations we have arbitrarily chosen to limit the production
to $100000$ modified mosquitoes per week. Some explorations have
managed to pass this limit. The present simulations always remain
below the limit. In practice, a smaller area requires fewer mosquitoes.
Therefore, the production limit is not an issue at this level. Repeatedly
hitting the limit would slow down or reduce the possibility of attaining
the targets of the treatment.
\item [{E.}] How does recycling of dead larvae and pupae affect the dynamics?
\\
Dead individuals could be reprocessed in different ways depending
on e.g., temperature, bacterial contents, etc. This issue is not intrinsic
to the RIDL-SIT technique apart from the fact that this or any other
technique must act on a given natural environment. The model allows
for sensing this issue to some extent. The effects appear to be less
important regarding the genetic diffusion but sensibly larger regarding
the temporary reduction in subpopulation sizes during the treatments.
\item [{F.}] Is it possible to eliminate the mosquito?\\
The model does not consider immigration (or emigration) to (from)
the intervention region. Under these conditions, the targeted treatment
will not eliminate the mosquito but rather replace it with some hybrid
strain. Simulations releasing weekly all the production of the factory
suggest that a $100\%$ Rockefeller strain with two copies of the
LG may replace the wild population (remaining stable) if the treatment
is long enough (see Appendix D). Analytical results confirm this observation.
Population levels are expected to be somewhat smaller and hence mosquitoes
larger than in the untreated situation.
\item [{G.}] How does the technique compare to other infrastructural measures
when it comes to costs, permanent benefits, risks and mosquito populations?\\
The model does not contemplate social and economic costs (in part
because of lack of information). The simulations suggest that the
technique will become a \emph{permanent patch}, this is, there will
be no end to the release programme and as such it represents a structural
cost that should be compared with other long-term policies (such as
running water, a sewage system, proper handling of water reservoirs,
etc.) that set the possibility of a healthier environment.
\end{lyxlist}

\section{Comparison with field observations\label{sec:field}}

Since the initial submission of the present work, two relevant articles
reporting on the evolution of \emph{Aedes aegypti} populations during
and after interventions have been published. It is relevant then to
compare quantitatively the field observations with our calculations.

The first work, \citep{garz17}, follows the release of genetically
modified mosquitoes in Jacobina and Juazeiro, two locations in Brazil.
The reported reduction in egg population is of $70$\%, and we will
compare their results with our runs for a $65$\% target (achieving
$68$\% reduction). Releases lasted for $10$ months (Juazeiro) and
$2$7 months (Jacobina). The authors report that reduction of the
population lasted for $17$ weeks and $32$ weeks after the end of
the releases. Actually, the data for Jacobina presents the first important
peak 14 weeks after the end of the releases. As the authors discuss,
these differences are expected because the times at which the intervention
ceased corresponds to different weather conditions for each case,
namely a favourable season (rainy, Juazeiro) and an unfavourable one
(dry, Jacobina). The present model for the $65$\% reduction target
estimates $13$-$18$ weeks for a full recovery of the population
under optimal weather conditions, the shorter times achieved when
$O1$ adults are considered unfitted (the results present little sensibility
to the fitness of $Rock$ hybrid type). In our model, a decrease of
90\% in the number of females carrying one or two LG's is observed
after $70$ days (\textasciitilde$2.5${} months) (for $O1$ unfitted
adults but fitted $Rock$ hybrids) while field data reports that ``transgenic
larvae were detected up to'' $2$ months (Jacobina) and $5$ months
(Juazeiro) after the release interruption. The recovery of the population
during the rainy (favourable) season observed in the field is somewhat
faster than in the model. The eradication of the population with LG
is produced by the competition with the wild type, a faster recovery
of the latter implies a faster eradication of larvae carrying the
LG, thus our overestimation of the time after the intervention when
they are detectable is linked to the same causes as the recovery time.
When $O1$ adults are considered fitted, the $O1$ eggs are no longer
present after $154$ days, a number harder to consider in agreement
with field data.

The second work, \citep{evan19}, measured hybridisation in Jacobina.
We will compare their results with a simulation for a $27$ months
intervention ($116$ weeks) at $65$\% reduction level (see Table
\ref{tab:Hybridisation-levels.-}). They report ``The degree of introgression
is not trivial. Depending on sample and criterion used to define unambiguous
introgression, from about $10$\% to $60$\% of all individuals have
some OX513A genome''. Their procedure is to compute an $h$-value,
namely the fraction of modified genes out of a set of $21770$ well
validated, biallelic, mendelian genes presenting polymorphism. $h$-values
larger than $0.02$ or $0.04$ are taken to be ``unambiguous introgressed
individuals''. The $h$-value may be compared with the average degree
of hybridisation $R$ in our simulations.

The $R$ value corresponds to an upper limit since the migration flux
favours a lower level of hybridisation measurable in the field and
the simulated intervention considers that the weather was favourable
to mosquito development all the time, but this is not true for the
case studies. Of the different cases of fitness of the hybrid population
studied, we can consider that the introgression simulated with $O1$
mosquitoes being as fit as wild mosquitoes produces an $R$ value
which is too high compared to field results. When mosquitoes carrying
the LG are considered unfitted, it is required that the $Rock$ mosquitoes
(hybrids with no LG) have a better fitness than the $O1$ in order
to match the observations. From our simulations and the reported field
studies it is not possible to conclude that the $Rock$ hybrids are
poor competitors in front of the Wild type.

In summary, the results are consistent with hybrids carrying the LG
being substantially unfit but first generations of $Rock$ hybrids
not being substantially different, in terms of fitness, to the Wild
type. At this point it is worth to consider Figure 2 in \citep{garz17}
where an increase in population is observed during the weeks $18$
(2013) and $16$ (2014) during the intervention time, and an increase
in population after week $16$ (2014), when the intervention had already
ended. This is, in the field data, the pre-intervention population
is lower than the post-intervention population. There are a number
of hypotheses that can account for such changes, and they are worth
to  be investigated, such as: Changing weather conditions, improved
genetics, changes in standard control measures applied and changes
in people's behaviour (relaxing environmental sanity).

Considering the limited information available about the local and
genetically modified populations, that the trials were performed in
a heterogeneous and open area in contrast with the model, and also
that weather data was not introduced in our simulations (and is not
available), the performance of our minimal-model must be considered
fairly good since it produces values in the same range as the corresponding
measurements for all the variables considered. 

\section{Final discussion and conclusions\label{sec:Conc}}

We have introduced a minimal model that allows study of hybridisation
produced by interventions with a particular form of genetically modified
mosquitoes. Several dynamically relevant factors had to be omitted
because of the lack of empirical information. In this respect, the
minimal model can serve as the basis of a more detailed model when,
and if, the information becomes available.

Introgression of genes (as an independent action) in a population
subject to natural selection results always in a population equally
or better adapted than the pre-existing population. This is a basic
theorem (supported on the mathematics behind genetic propagation)
in as much as fitness can be associated with the genetics of the population.
Only the quality and quantity of the improvement remains to be ascertained
in each situation. The same theorem indicates that the result of lowering
genetic variability will in general decrease fitness, the limiting
case being a null decrease. Hence, if controlling the population with
a SIT technique results in a decrease of variability, the total outcome
of the RIDL-SIT experiment is completely uncertain.

When results from the model are compared to recent interventions we
observe that they are quantitatively compatible when hybrids carrying
the LG are considered to have a mating efficiency equal to mosquitoes
carrying two copies of the LG and an increased mortality rate, while
hybrids carrying no LG are closer in mating and survival to the wild
population. This situation implies that if $F1$ males (all of them
pertaining to the $O1$ class of the model) are as unfitted as released
males and $F1$ females (and all members of the $O1$ class) die twice
as faster that wild females (an exaggeration), genetic diffusion still
reaches important levels. The fitness relative to the Wild population
cannot be asserted by the present methods. The intervention is not
self-cleaning and will leave behind genetic contamination with the
unavoidable consequences stated by the theorem just mentioned.

The present model is a ``proof of concept'' rather than a ``production
model''. One of the goals of this work has been to show how to model
the problem of genetic diffusion and to highlight its relevance. Mosquito
models can be highly elaborated and as such they are sensible to local
differences in populations or environment. A compartmental model intending
to forecast the outcome of an intervention can be conceived by merging
the modelling strategies developed in this work into a detailed model.
The information needed to asses particular interventions includes
the reaction norms for developmental treats as a function of genetics
(wild (local) mosquitoes, $F1$ hybrids, $F1b$ hybrids), concentration
of food and temperature. The mating efficiency of $F1$ and $F1b$
males is needed as well (along with more standard information such
as several years of weather records: at least rain and average daily
temperature).

The processes of development of models and of introduction of new
technologies present some parallels that are worth discussing. In
the first place, both processes respond to the different attitudes
towards knowledge and learning, or more generally to the epistemic
frame(s) in action.

From pre-modelling (intuition-based) conclusions we enter the realm
of theoretical conclusions based on mathematical models. A first stage
in model development can be called the \emph{chimeric level} where
the modelled mosquito only roughly resembles the real one (take this
expression as self-criticism \citep{oter06,oter08}), but at least
the time-scales and life-cycle characteristics are present. Models
mature into ecologically oriented models \citep{rome13,rome15,rome15b}
fundamentally including phenotypic plasticity \citep{reed10} and
focusing on one strain at a time (subpopulations of mosquitoes in
different cities may be strikingly different \citep{Tejerina2009,Grech2010}
and even within the same city \citep{paup04}, different genetics
is associated to different breeding sites) and may further progress
towards evolutionary models \citep{sche93} accounting for the evolution
of phenotypic plasticity. In this higher level we still have everything
to learn.

In parallel, technologies start as laboratory tests, essentially ignorant
of the hazards and singularities of life outside the laboratory. At
this level, technologies are short-sighted; they cannot look ahead
the time and space scales of the controlled laboratory tests. In order
to mature into reliable tools, they too have to develop, including
insight about the mutual influences between technologies and environment.
This triggers the need of a research programme and an action policy
which eventually may outcome a strategy to handle the challenges of
the environment and of (co)evolution.

The perspectives of the RIDL-SIT technique in \emph{Aedes aegypti}
considered before this work are clearly at the first stage. We have
pointed out in the previous sections issues about missing information
on both the life-cycle of the modified mosquitoes and on their interaction
with the environment and with the wild strain. Moreover, already at
the level of bibliographical background research we note that important
requirements mentioned e.g., in \citep{bene03} are not fulfilled.

The information so far produced corresponds to the development of
the technology and extrapolations of laboratory results based upon
intuitions. We have termed this epistemic frame the \emph{technological
frame} \citep{sola16b,sola17}. This issue has been discussed as:
``typological versus population thinking in biology`` as well \citep{powe18}.
Typological thinking usually implies the fundamental decision of not
exploring the differences between a wild environment and the laboratory.

In this work we implemented an environment-aware model of the life-cycle
of a strain of \emph{Aedes aegypti} compatible with field observations
\citep{rome13,rome15,rome15b} also introducing some genetics since
the issue at stake with the RIDL-SIT technique is to take advantage
of genetic differences between mosquito strains. We have demonstrated
that, with the help of mathematical models, it is possible to detect
some of the information needed to make less risky decisions than just
``trial and error'' (which has been the method of technology in
all civilisations but it is not the method of science), and that there
is more to monitor than the success of the method considered only
in the restricted terms of the developers of the technology. The model
suggests that not even in a minimal area protected from immigration
of wild specimens the technique is able to eliminate the mosquito
population. Even worse, the output of the treatment is to produce
a new, hybrid, mosquito strain, whose characteristics are unknown.
The RIDL-SIT strategy can be viewed as a technological patch for an
unsustainable control programme, a conclusion that has been reached
previously from social considerations \citep{reis13}. Indeed, the
results of this work indicate that there is a long path that needs
to be travelled before moving into field tests.

The limited nature of the model manifests the need of understanding
evolutionary issues in order to proceed further: How might the mosquito
change under selective pressure? How will the differences between
hybrid and wild mosquitoes manifest themselves? The results of hybridisation
cannot be forecast in the present context of lack of information,
yet this difficulty should not be rephrased as ``there are no consequences
expected from hybridisation''. In all cases, the larger genetic pool
available will result in a better matching of the local forms of the
mosquito to the environment. This situation opens a large range of
possibilities we only can speculate about. They go from a return to
the pre-release equivalent epidemic risk to situations of increased
risk. Let us consider two possible mechanisms for the latter. If the
larger fertility provided by the Rockefeller background of the released
mosquitoes results in larger fertility of the new hybrid population
established in place of the original wild population, without altering
the available biomass of food, a larger number of adults of smaller
size than the original wild population is to be expected. A larger
number of vectors facilitates the propagation of mosquito transmitted
diseases, while the favourable or unfavourable incidence of the size
of the vector is a matter of discussion \citep{juli14}. In the same
form, a co-evolution of bacteria in which \emph{Ae ae.} feeds prompted
by the increase in mortality in the breeding sites is expected to
result in substantially larger populations because of the sensitivity
detected to ``recycling'' dead pupae and larvae. Thus, an ``arms
race'' scenario might arise (increased population retaliated by increased
releases). The epidemiological situation after the intervention could
then be worse than before the intervention. Some detailed models like
SkeeterBuster predict overshooting of mosquito populations going above
the equilibrium value after interventions, others like aedesBA do
not \citep{legr16}, it is a matter under discussion that depends
on modelling decisions under uncertainty. We must conclude that the
possibility of a renewed risk for epidemics after the intervention
has ceased should be taken into account. The observed increases in
population during and after the Juazeiro tests should be addressed,
they cannot be ignored.

This work, as much as previous experience in modelling \emph{Aedes
aegypti} populations, indicates that local differences in environment
and mosquito's phenotype are reflected in the population dynamics
in sensible forms. This observation warns us about the risks of attributing
to the species what is in fact the result of the genetics and environment.
Mosquito strains are different, they interact in (so far) unexpected
ways and are subject to a large, and mostly unknown today, number
of influences. The technological frame (or typological thinking) excels
at the time of proposing interventions but ecological or population
thinking must be used in evaluating the proposals. In the same direction,
it would be a mistake to extrapolate the situation studied in this
work to other conceivable cases of RIDL-SIT; the evaluation of every
case must be performed based on actual measurements of biological
traits, and such experimental values should be produced. Possible
progress is then conceived as a dialogue between both biological views.
The present work indicates that mathematical modelling can be an important
form to explore proposals and possible outcomes.

The standard analogy between controlling fruit flies and controlling
mosquitoes such as \emph{Ae ae.} is not a very close one from an ecological
point of view. A very obvious difference is the landscape, going from
agricultural to urban, yet a deeper difference consists in that \emph{Ae
ae.} is an established domestic insect limited only by the environment,
while the fruit fly, when production has not been abandoned, is not
limited by the available environment, but rather the fly is just attempting
to colonise it, being in short number relative to the carrying capacity.
Current knowledge in SIT indicates that ``releasing sterile insects
routinely in certain areas may be more expedient to prevent establishment
of major pests than eliminating them after they become established''
\citep{hend05}.

\section*{Acknowledgements}

HGS acknowledges support from the University of Buenos Aires under
grants 20020130100778BA and 20020130100361BA, as well as support from
the MOSTICAW (STIC-AMSUD) collaboration. MAN acknowledges support
from SveFUM and from Kungliga Fysiografiska Sällskapet i Lund.


\newpage{}

\appendix

\section{Lethal-gene enhanced mortality at pupae level }

The assumptions behind the modified death and emergence rates for
pupae are as follows. For the model, pupae can undergo only two events,
namely emergence $m_{p\to a}$ and death $m_{p}$. We assume that
the time to the next event is not altered with respect to the wild
population, so in both cases this time responds to an exponential
distribution with parameter $R_{pupa}=m_{p\to a}+m_{p}=m_{pl\to a}+m_{pl}$
(where $l$ stands for ``lethal''). Further, we assume that $5\%$
of pupae carrying the LG emerge while the other $95\%$ dies. Hence,
\[
\frac{m_{pl\to a}}{m_{pl\to a}+m_{pl}}=0.05
\]
 and therefore
\[
\frac{m_{pl\to a}+m_{pl}}{m_{pl\to a}}=20\quad\mathrm{or}\quad m_{pl}=19m_{pl\to a}
\]
Hence,
\begin{eqnarray*}
m_{pl\to a} & = & \frac{1}{20}\left(m_{p\to a}+m_{p}\right),\quad\mathrm{and}\\
m_{pl} & = & \frac{19}{20}\left(m_{p\to a}+m_{p}\right).
\end{eqnarray*}

\section{Deterministic equilibrium population}

Deterministic population equations as those are frequently seen in
epidemiological modelling, have an older history \citep{verh38} than
the stochastic approach. However, deterministic models have a more
limited value, since they average away relevant issues of the dynamics.
The most conspicuous limitation is that these models cannot properly
describe few individuals, as would be the case in problems where there
is a potential extinction risk (such as ours). In certain situations
\citep{ethi86,sola03} the deterministic dynamics can be seen as a
sort of large-population limit of the stochastic approach. However,
since the wild population is not spontaneously facing extinction,
the deterministic equilibrium values are a reasonable starting point
for our simulations.

To produce deterministic equations out of the stochastic setup amounts
to combine the rates $W_{\alpha}$ and associated modifications $\delta_{j}^{\alpha}$
populationwise.

The general expression for the deterministic dynamics of each population
species in the absence of released individuals reads, 
\[
\frac{dX_{j}}{dt}=\sum_{\alpha}W_{\alpha}(X_{1},\cdots,X_{n})\delta_{j}^{\alpha}=\sum_{\alpha}m_{\alpha}^{i}X_{i}\delta_{j}^{\alpha}
\]
where the sum over $\alpha$ is relevant only for those events modifying
population $X_{j}$ and $X_{i}$ is the specific subpopulation associated
to the event with coefficient $m_{\alpha}^{i}$. At this point recall
that fecundation is not treated as an event, i.e., it is assumed to
occur in a time-scale that is negligible in front of the life-span
of adult females. Letting $E,\,L,\,P,\,F$ denote the size of the
wild populations of eggs, larvae, pupae and females, we obtain, 

\begin{eqnarray*}
\frac{dE}{dt} & = & m_{ovi}\,c_{lay}FT(\frac{P_{f}}{C_{f}})\,F-(m_{e\to l}+m_{e})E\\
\frac{dL}{dt} & = & m_{e\to l}E-LE(\frac{P_{f}}{C_{f}})\,L\\
\frac{dP}{dt} & = & LE(\frac{P_{f}}{C_{f}})\,\left(1-ML(\frac{P_{f}}{C_{f}})\right)\,L-0.5787\,P\\
\frac{dF}{dt} & = & \frac{1}{2}0.5787\,\left(1-ML(\frac{P_{f}}{C_{f}})\right)P-\frac{0.04}{1-ML(\frac{P_{f}}{C_{f}})}F
\end{eqnarray*}
The food equilibrium condition influences all equations and also the
larvae population $L$. Being the deterministic system quasilinear
\citep{sola14} in the populations, the equilibrium is obtained when
the determinant of the coefficient matrix is zero, determining in
such a form the food level of the equilibrium. In terms of exponential
races, the equilibrium is independent of the average time required
for the race and depends only on the probability of occurrence of
one or another outcome. The population number is determined by the
production of food by the environment. The equilibrium values were
obtained solving the equilibrium food dynamics to obtain the equilibrium
$\frac{P_{f}}{C_{f}}$ value and the equilibrium larvae population
$L$. Subsequently, all other populations were obtained by solving
$\frac{dE}{dt}=0,\,\frac{dP}{dt}=0,\,\frac{dF}{dt}=0$ as a function
of $L$ and $\frac{P_{f}}{C_{f}}$ . 

\section{Reaction norms determined for this work}

\begin{figure}[h]
\centerline{\includegraphics[bb=0bp 0bp 665bp 465bp,width=6cm,height=4cm]{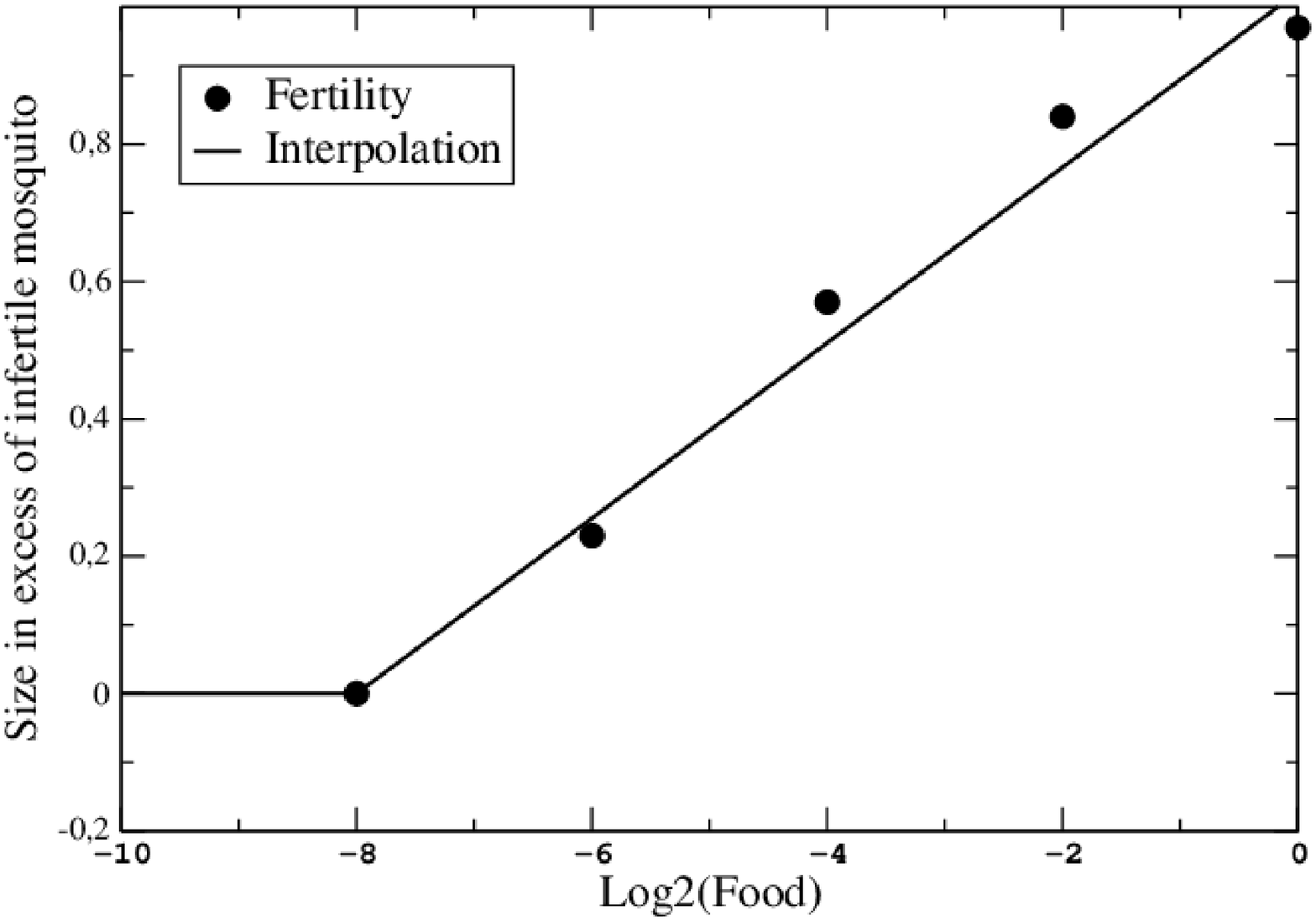}\includegraphics[width=6cm,height=4cm]{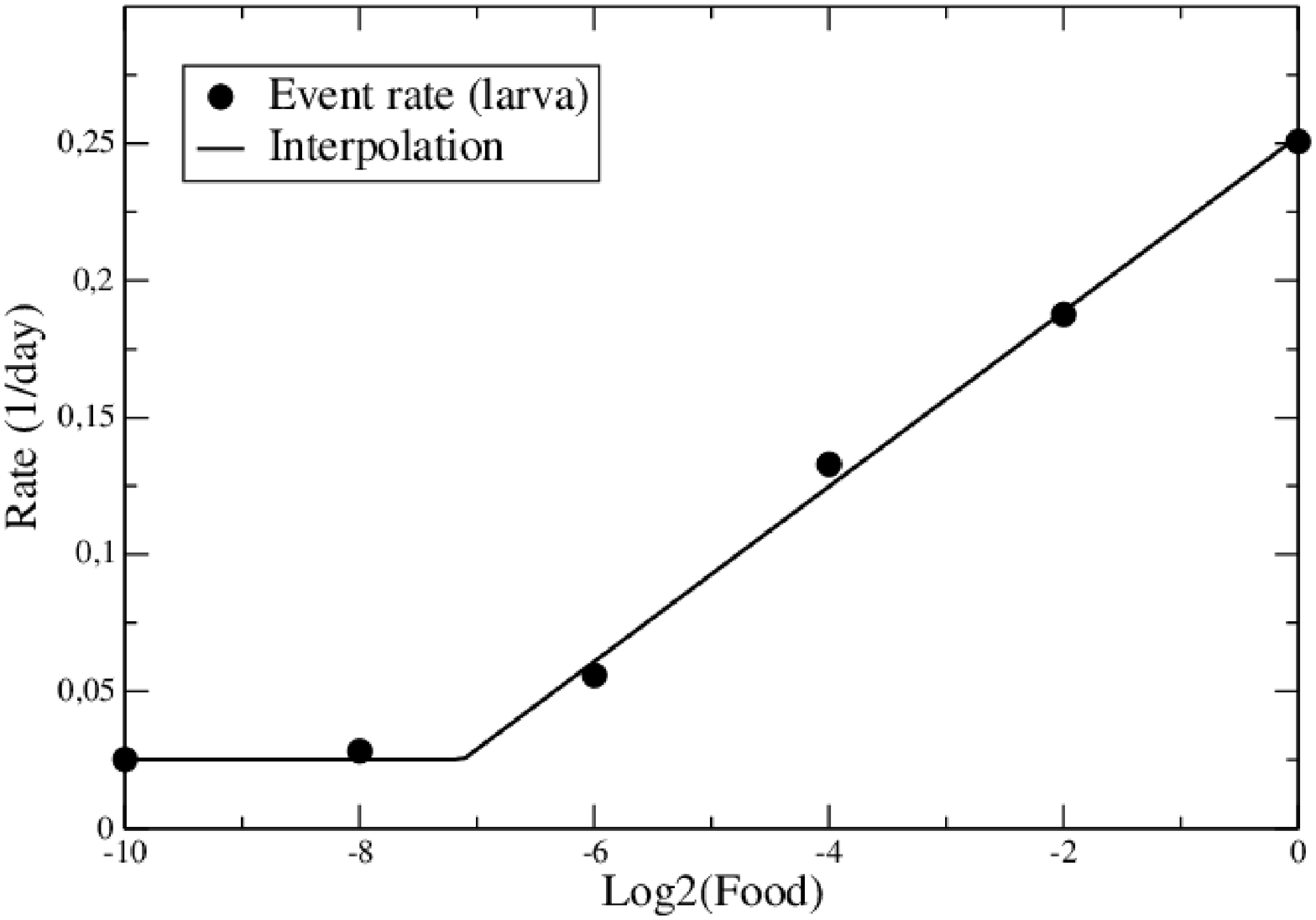}}

\caption{Reaction norms for fertility and event rate (death plus pupation)
for larvae. (No females emerged for the $2^{-10}$ treatment).\label{reaction-norm}}
\end{figure}

The data available from the experiment reported in \citep{rome15,rome15b}
was used to produce the estimates for the dependence of body size
(surrogated by wing length) and the rate for events (pupation or death)
as larvae. The results are displayed in Figure \ref{reaction-norm}.
The data displays clearly two different responses to food depletion:
the starvation region where the probability of mortality increases
but the average time for the next event in the larvae compartment
does not change significantly; and the scarcity region where mortality
is not increased but the average time between events (the reciprocal
of the rate) changes monotonically. The left panel of Figure \ref{reaction-norm}
shows the wing length as a function of the logarithm of the fraction
of optimal food density. Wing length, body size and fertility are
monotonically related. The right panel shows the event rate for larvae
as a function of the same environmental variable.

\section{Constant vs targeted release}

The goal of control strategies is to reduce mosquito populations to
a level in which epidemic outbreaks are not possible. Releasing genetically
modified mosquitoes in excess of those needed to reach the target
represents a misuse of costly resources. A compromise must be taken
between the area under control and the degree of control (meaning
the number of mosquitoes per human being for example).

\begin{figure}[h]
\centerline{\includegraphics[width=7.5cm]{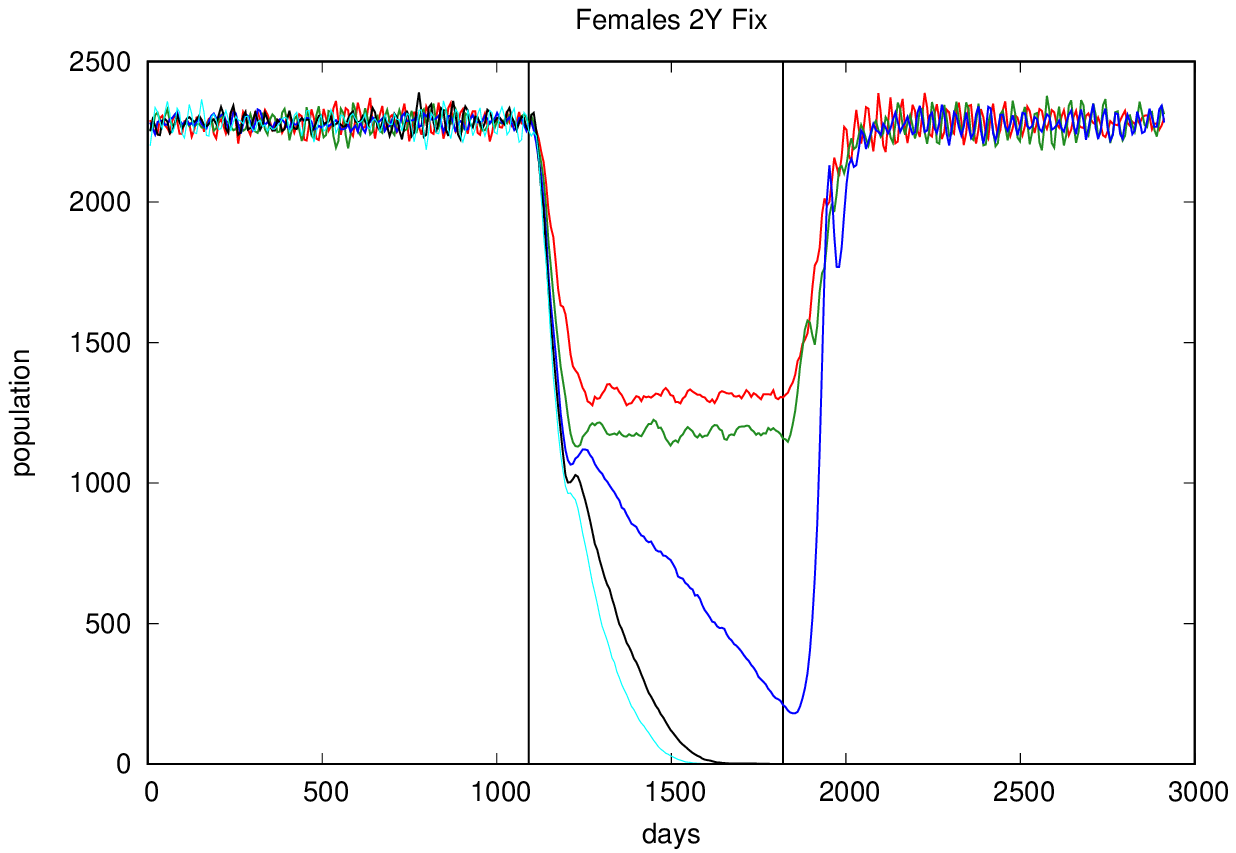}\includegraphics[width=7.5cm]{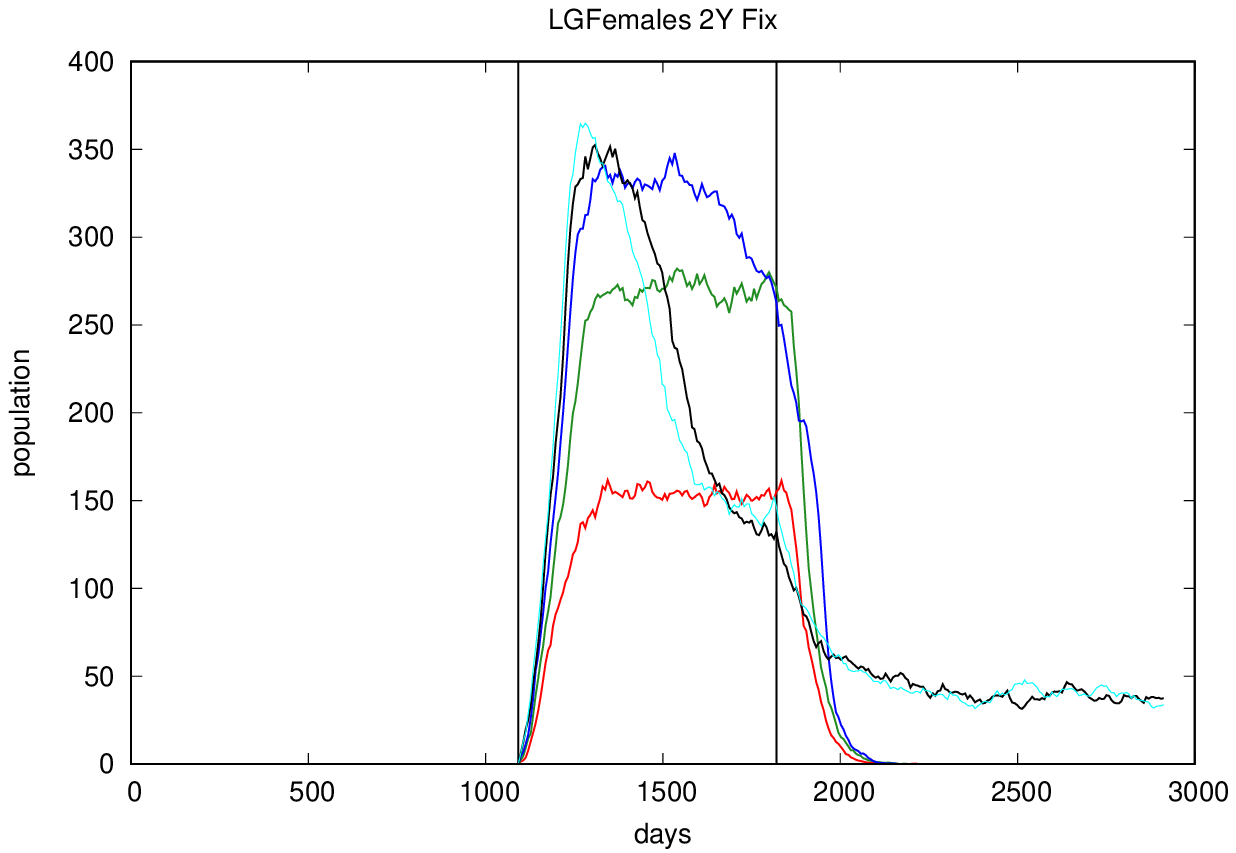}}

\caption{Females with no LG (left) and with 2 LG (right) for constant weekly
release of 10000 (red) to 50000 (light blue) individuals (green: 20000,
dark blue 30000, black 40000 individuals).\label{fig:const-rel}}
\end{figure}

In contrast, in a trial phase, one may want to test how to achieve
maximal effect. To depict this situation we simulated five ``large''
constant weekly releases (larger or much larger than the maintenance
release targeted in Figures \ref{fig:Females1}, \ref{fig:Females5}
and \ref{fig:graph3}), acting under two years, for the case of 12\%
food recycling. Only for the larger values the wild and hybrid (with
no LG) populations are outnumbered to extinction in the present modelling
conditions (closed environment with no immigration/emigration of individuals,
no heterogeneities), while a smaller population of 100\% Rockefeller
with 2 LG individuals persists for over three years after the intervention.
Results are displayed in Figure \ref{fig:const-rel}. However, this
situation is unrealistic in field conditions where immigration/emigration
cannot be controlled.

\end{document}